\documentclass[12pt]{article}
\usepackage[utf8]{inputenc}
\usepackage{graphicx,psfrag,epsf}
\usepackage{booktabs}
\usepackage{textgreek}
\usepackage{threeparttable}
\usepackage{float}
\usepackage{amsmath,amsfonts,amsthm,bm} % Math packages
\usepackage{url}
\usepackage{transparent}
\usepackage{adjustbox}
\usepackage{flafter}
\usepackage{lscape}
\usepackage{caption}
\usepackage{subcaption}
\usepackage{graphicx}
\usepackage{geometry}
\usepackage{footnote} % For footnotes in a table
\makesavenoteenv{tabular} % For footnotes in table
\usepackage{verbatim}
\usepackage{natbib}
\usepackage{comment}
\usepackage{rotating}
\usepackage{hyperref}
\usepackage{mdframed}
\usepackage{lipsum}
\usepackage{array}
\newcolumntype{H}{>{\setbox0=\hbox\bgroup}c<{\egroup}@{}}
\usepackage{xcolor}
\usepackage{amsmath}
\usepackage{multirow}
\usepackage{bbm}
\usepackage{algorithmicx}
\usepackage{algorithm,algpseudocode}
\usepackage{longtable}

\usepackage[final, commandnameprefix=ifneeded]{changes}
\newcommand{\blind}{0}

\begin{document}

\def\spacingset#1{\renewcommand{\baselinestretch}%
{#1}\small\normalsize} \spacingset{1}

%----------------------------------------------------------------------------------------
%	Title Page Section
%----------------------------------------------------------------------------------------
\def\spacingset#1{\renewcommand{\baselinestretch}%
{#1}\small\normalsize} \spacingset{1}

%%%%%%%%%%%%%%%%%%%%%%%%%%%%%%%%%%%%%%%%%%%%%%%%%%%%%%%%%%%%%%%%%%%%%%%%%%%%%%

\if0\blind
{
  \title{\bf Crossing penalised CAViaR\thanks{
    Special thanks go to Mark Schaffer for discussions that were crucial in shaping the methodology. Further thanks go to Arnab Bhattacharjee, Rod McCrorie, and Atanas Christev for their comments on early versions of the paper. I would also like to thank the ESRC for PhD studentship as well as Heriot-Watt University for institutional support. The usual disclaimer applies.}}
  \author{Tibor Szendrei \footnote{Corresponding author: t.szendrei@niesr.ac.uk.}\\
  National Institute of Economic and Social Research, UK \\
    Edinburgh Business School, Heriot-Watt University, UK %\\
    % \\
    %Mark E. Schaffer\\
    %Edinburgh Business School, Heriot-Watt University, UK
    }
  \maketitle
} \fi

\if1\blind
{
  \bigskip
  \bigskip
  \bigskip
  \begin{center}
    {\LARGE\bf Title}
\end{center}
  \medskip
} \fi

\newtheorem{assumption}{Assumption}

%----------------------------------------------------------------------------------------
%	Abstract Section
%----------------------------------------------------------------------------------------
%\begin{center}
%    \textbf{Early Draft! Please do not disseminate!}
%\end{center}

\bigskip
\begin{abstract}
\noindent Dynamic quantiles, or Conditional Autoregressive Value at Risk (CAViaR) models, have been extensively studied at the individual level. However, efforts to estimate multiple dynamic quantiles jointly have been limited. Existing approaches either sequentially estimate fitted quantiles or impose restrictive assumptions on the data generating process. This paper fills this gap by proposing an objective function for the joint estimation of all quantiles, introducing a crossing penalty to guide the process. Monte Carlo experiments and an empirical application on the FTSE100 validate the effectiveness of the method, offering a flexible and robust approach to modelling multiple dynamic quantiles in time-series data.
\end{abstract}

%----------------------------------------------------------------------------------------
%	Key words
%----------------------------------------------------------------------------------------

\noindent%
\noindent
%{\it JEL:} 
\vfill

\spacingset{1.45} % DON'T change the spacing!

%----------------------------------------------------------------------------------------

\section{Introduction}
The idea of quantiles depending on past values of the quantiles themselves has been popularised in the Value-at-Risk literature. These processes are often referred to as Conditional Autoregressive Value-at-Risk (CAViaR) \citep{engle2004caviar} and have been used to estimate single quantiles. This common approach has limitations when one is interested in estimating the parameters for multiple quantiles. In particular, one often ends up in situations where the estimated quantiles cross. This is a problem as we would like estimated quantiles to behave like true quantiles, i.e., they should monotonically increase.

The problem of quantile crossing motivated \citet{gourieroux2008dynamic} to develop a dynamic quantile model that yields non-crossing quantiles. The method remedies the problem of crossing quantiles by introducing path dependent parameters. Furthermore, the method nests nicely within the GARCH framework often used to estimate VaRs. However the estimation method is not semi-parametric in nature, unlike the regression quantiles of \citet{koenker1978regression}.%, and therefore limited in modeling nonlinearities.

This paper develops a method that can be used to obtain non-crossing CAViaR curves without the need to specify the error process. This is achieved by adopting a penalised quantile regression framework. The usual linear programming setup cannot be used for estimation here because the lagged quantiles are unobserved by default. To estimate this model we use a general purpose solver, following \citet{engle2004caviar}, \citet{white2010modeling} and \citet{white2015var}. However, unlike the above papers, we use the covariance matrix adaptation evolution strategy (CMA-ES) of \citet{hansen2009tec} to minimise the function, rather than the Nelder-Mead algorithm, due to the ``rugged'' nature of the objective function.\footnote{An objective function is defined as ``rugged'' if it possesses many locally optimal solutions \citep{hansen2009tec}.}

%\newpage
To evaluate the performance of the method, we conduct a small Monte Carlo experiment to compare two processes: QAR (quantile autoregressive) models of \citet{koenker2006quantile}; and CAViaR of \citet{engle2004caviar}. QAR processes can be estimated using standard quantile regression methods and serve as a benchmark for our proposed method. %To generate the processes, we use the method outlined in \citet{white2010modeling}. 
The Monte Carlo experiments reveal that the proposed methodology is capable of providing estimates with lower coefficient bias for the different DGP's generated. %Furthermore, as the hyperparameter regulating the penalty on crossing is increased, the probability of crossing decreases. 
Importantly, the choice of initial conditions has limited impact on the estimation. Furthermore, in addition to lower coefficient bias, the proposed method also yields lower probability of quantile crossing as compared with the estimator of \cite{engle2004caviar} and \citet{white2010modeling}.

This study investigates the performance of various quantile regression estimators for modelling and forecasting the volatility of the FTSE100 index during the onset of the Global Financial Crisis. The year 2008 was chosen due to the extreme market fluctuations and increased volatility, offering a challenging environment to assess the effectiveness of different quantile-based approaches. We compare the performance of models that incorporate quantile inertia with more traditional quantile regression estimators. By analysing both in-sample fits and out-of-sample forecast accuracy, we aim to highlight the strengths and weaknesses of these estimators in capturing asymmetric risk dynamics and the persistence of volatility during periods of financial stress. 

Our results demonstrate that quantile models that incorporate lagged quantiles significantly outperform traditional quantile regression methods in capturing the evolving volatility of the FTSE100 index during the financial crisis of 2008. These dynamic models, which incorporate lagged quantiles, provide smoother, more accurate forecasts, especially for the tails of the distribution. In contrast, traditional methods struggled to adjust to extreme market shifts, often overshooting or undershooting during periods of heightened risk. The proposed DynQR estimator, with $\lambda>0$, showed superior forecast accuracy highlighting its robustness in forecasting under conditions of increased market uncertainty. Furthermore, the proposed method showcased more stable lagged quantile coefficients than the CAViaR-NM proposed in \citet{white2010modeling}. These findings suggest that dynamic quantile models that penalise quantile crossing offer a more reliable approach for forecasting in volatile financial markets, with important implications for risk management strategies.

%Our empirical application is to Growth-at-Risk (GaR) popularised by \citet{adrian2019vulnerable}. This paper constitutes the first time a CAViaR like method is used to estimate GaR. One finding of \citet{adrian2019vulnerable} is that upper quantiles are more stable than lower quantiles. This implies that upper quantiles have higher lagged quantile coefficient than lower quantiles. We test this hypothesis using the proposed method. Our US GaR application shows that the upper quantiles are characterised by ``sticky'' quantiles. This in turn leads to better right tail forecasts compared to traditional QR estimators. 

The paper is structured as follows. Section~2 introduces the methodology, focusing on the objective function as well as the estimator employed. The Monte Carlo setup is described in Section~3 followed by discussion of coefficient bias of the different estimators. The Monte Carlo section also considers computation efficiency, particularly the average time it takes to obtain a solution utilising a single core. In Section~4, the method is applied to FTSE100 during 2008. Finally, Section~5 concludes.

\section{Methodology}
\subsection{Assumptions}
Just like in \citet{white2010modeling}, the objective of this paper is to estimate several quantiles jointly. To this end, we adopt several of the assumptions made in \citet{white2010modeling}.%, in particular:

\begin{assumption}
\textit{Stationarity and Ergodicity:} The sequence ($Y_t$,$X_t^T$) is a stationary and ergodic stochastic process. $Y_t$ is a scalar and $X_t$ is a vector with its first element being 1. The probability space for the sequence is complete and defined as ($\Omega$,$\mathcal{F}$,$P_0$).
\end{assumption}

Let $\mathcal{F}_{t-1}$ be the $\sigma$-algebra generated by $Z_{t-1}=\{X_t,(Y_{t-1},X_{t-1}),...\}$. Define the cumulative distribution function of $Y_t$ conditional on $\mathcal{F}_{t-1}$ as $F_t(y)=P_0[Y_t<y|\mathcal{F}_{t-1}]$. Further, let $\tau\in(0,1)$ be a series of quantiles of interest, with $Q$ monotonically increasing elements such that $\tau=\{ \tau_1,\tau_2,...,\tau_Q \}$. Given the CDF and the quantiles we can define the $\tau_q^{th}$ quantile of $Y_t$ conditional on $\mathcal{F}_{t-1}$, which we denote with $\hat{\mathcal{Q}}_{\tau_q,t}$, as:

\begin{equation}
    \hat{\mathcal{Q}}_{\tau_q,t}=F_t^{-1}(\tau_q).
\end{equation}

\begin{assumption}
\textit{Continuity of Density:} $Y_t$ is continuously distributed for each $t$ given the parameters $\omega\in\Omega$, such that the CDF ($F_t$) and the PDF ($f_t$) are continuous on $\mathbb{R}$. Suppose that $f_t(\omega,\mathcal{Q}_{\tau_q,t})>0$. Further assume that $x_t$ is a $(K+1) \times 1$ stationary and ergodic sequence of random vectors.
\end{assumption}

Given these assumptions we can represent a data generating process that is a multi-quantile version of the CAViaR introduced by \citet{engle2004caviar} \citep{white2010modeling}. The multi-quantile CAViaR is defined as:

\begin{equation}
    \mathcal{Q}_{\tau_q,t}=x_t^T\beta_{\tau_q}+\sum^L_{l=1}\mathcal{Q}_{\tau_q,t-l}^T\theta_{\tau_q,l}.
\end{equation}

There are two types of coefficients in the above formulation: the $\beta$ parameters associated with the covariates, which can contain lags of $Y_t$, and the $\theta$ parameter that determines the degree of inertia in the evolution of the quantiles. The coefficient vector, that we wish to estimate, is defined as $\delta^T=(\beta^T,\theta^T)$ which has dimensions $\ell \times 1$, where $\ell=Q(K+1+LQ)$.

\begin{assumption}
\textit{Continuity of coefficient vector:} Let $\mathbb{D}$ be a compact subset of $\mathbb{R}^\ell$. The sequence of $\mathcal{Q}_t(\omega,\delta)$: (1) is measurable for each $t$ and $\delta \in \mathbb{D}$, and; (2) is continuous for each $t$ and $\omega\in\Omega$ on $\mathbb{D}$.
\end{assumption}

The assumption about the continuity of the coeffcient vector is particularly crucial as discontinuous functions can lead to difficulties in optimisation, making it hard to find the parameter estimates that minimise the objective function.

\begin{assumption}
    \textit{Correct Specification and Identification:} (1) There exists $\delta^*\in\mathbb{D}$ such that $P[Y_t\leq\mathcal{Q}_{\tau_q,t}(\cdot,\delta^*)]=\tau_q~\forall t,q$. (2) For some $\psi>0$, with $||\delta-\delta^*||\geq\psi$, and $\delta\in\mathbb{D}$, we have $P[Y_t\leq\mathcal{Q}_{\tau_q,t}(\cdot,\delta)]\neq\tau_q$.
\end{assumption}

This assumption ensures that the true parameter $\delta^*$ lies in the parameter space $\mathbb{D}$ and correctly specifies the quantiles. The second part of the assumption ensures that $\delta^*$ is unique. Without this assumption, there could be multiple sets of parameters that fit the data equally well, which would make optimisation difficult.

%\newpage
%Define the cumulative distribution function of $Y_t$ as $F_t(y)=P_0[Y_t<y|\mathcal{F}_{t-1}]$. Further, let $\tau_q\in(0,1)$ be a series of quantiles of interest that are monotonically increasing. Given the CDF and the quantiles we can define the $\tau_q^{th}$ quantile of $Y_t$ conditional on $\mathcal{F}_{t-1}$, which we denote by $\mathcal{Q}_{\tau_q,t}$:

%\begin{equation}
%    \mathcal{Q}_{\tau_q,t}=F_t^{-1}(\tau_q).
%\end{equation}

%Just like in \citet{white2010modeling}, the objective of this chapter is to estimate several quantiles jointly: $\tau=\{ \tau_1,\tau_2,...,\tau_Q \}$. 
We deviate from \citet{white2010modeling} in two ways. The first change is that we amend the objective function to reflect the findings of \citet{bondell2010noncrossing}, who note that simply jointly estimating quantiles does not necessarily yield sufficiently different results from estimating separate quantiles. To obtain different results one needs to introduce some cross-quantile constraints. In this paper, these cross-quantile constraints are implemented via a penalised regression framework. Specifically, we penalise quantile crossing, and thereby propose a non-crossing CAViaR estimator. 

We note that jointly estimating a CAViaR structure could potentially yield different results if the different unobserved quantiles have significant impacts upon each other. Nonetheless, in \citet{white2010modeling}, the cross quantile coefficients are near 0. Accordingly, in this paper we restrict the cross-quantile effects to 0. Note that this simplifying assumption can be relaxed and the corresponding changes to the objective function are trivial.

Our second modification lies in the optimisation routine utilised. In \citet{white2010modeling}, the individual CAViaR models are first estimated individually using a series of Nelder-Mead algorithm (hereinafter NM)\footnote{For more information on the optimisation routine see \citet{nelder1965simplex}.} iterated until convergence. Then, using the quantile specific parameter estimates as an initial point, the same series of calculations are done for the joint estimation routine. Rather than executing this series of routines, we use the CMA-ES algorithm of \citet{hansen2009tec}, which handles ``rugged'' search spaces better.

\newpage
\subsection{Objective function}
Our starting point for the objective function is the CAViaR of \citet{engle2004caviar}:

\begin{equation} \label{eq:CAViaR}
    \hat{\mathcal{Q}}_{\tau_q,t}=x_t^T\beta_{\tau_q}+\sum^L_{l=1}\hat{\mathcal{Q}}_{\tau_q,t-l}^T\theta_{\tau_q,l}+\varepsilon_t.
\end{equation}

When $L=0$, the above equation is simply the quantile regression estimator of \citet{koenker1978regression}. By allowing $L>0$, we start introducing quantile ``stickiness'', in the sense that the given quantile cannot change from one period to the next as freely as it would in conventional quantile regression. This approach results in volatility clustering which is of key interest in finance and macroeconomics and has been traditionally modelled by GARCH models of \citet{bollerslev1986generalized}. Thus, our work is also related to the literature on extending the GARCH model to quantile regression; see, for example, \citet{koenker1996conditional}, \citet{engle2004caviar} and \citet{xiao2009conditional}. Throughout this paper we assume that $L=1$.

Quantile crossing has been a problem in quantile regression and several solutions have been proposed \citep{he1997quantile, bondell2010noncrossing,chernozhukov2010quantile}. The quantile crossing problem can be more pronounced in CAViaR type models as the different quantiles could potentially have different degrees of inertia, which would lead to situations where the more ``agile'' quantiles cross the more ``sticky'' ones. While the method of \citet{chernozhukov2010quantile} is applicable here, if the interest lies on inference on the estimated coefficients, one needs to impose constraints in the estimation \citep{bondell2010noncrossing}.

Imposing non-crossing constraints is attractive, but implementation is not straightforward in the CAViaR setting. In particular, the past values of the quantile are not observed and as such the transformation from \citet{bondell2010noncrossing} cannot be implemented. \citet{szendrei2023fused} show that non-crossing constraints are just a special quantile specific fused shrinkage, but their proposed method also cannot be implemented as easily for the CAViaR setting. In particular, without rescaling the data, one needs quantile specific hyperparameters, which would entail increasing the dimension of the grid and as such increase computation time exponentially. As such, we make adjustments to the multi-quantile CAViaR objective function by adding explicit non-crossing constraints. Consider the canonical constrained optimisation setting:
\begin{equation} \label{eq:constrainedoptim}
    \begin{split}
        \hat{\delta}=\underset{\delta}{argmin}&\frac{1}{QT}\sum^{Q}_{q=1}\sum^{T}_{t=1}\rho_{\tau_q}(y_{t}-x_t^T\beta_{\tau_q}-\hat{\mathcal{Q}}_{\tau_q,t-1}^T\theta_{\tau_q})\\
        s.t.~&\hat{\mathcal{Q}}_{\tau_q,t}\geq\hat{\mathcal{Q}}_{\tau_{q-1},t},
    \end{split}
\end{equation}
where $\hat{\mathcal{Q}}_{\tau_q,t}$ is the fitted quantile in the form of equation (\ref{eq:CAViaR}). While intuitively appealing, this formulation needs imposing $(Q-1)\times T$ constraints, which is untenable for computations. To address this issue, \citet{bondell2010noncrossing} rescaled the problem to the domain of $[0,1]$. This option is not available here because the quantile lags are not observed. Hence, we need an explicit measure of quantile crossing in the constraint.

This raises the question as to how to measure quantile crossing. A simple approach is to compare the estimated and sorted quantiles and count the instances where the two vectors are not equal. While conceptually simple, the problem is that this measure is discrete, and introducing it in the objective function can cause challenges for optimisation. Instead, we opt to include a direct measure of crossing. The most obvious measure of crossing is to look at the fitted quantiles and count the number of crossing quantiles, i.e. $\sum^T_{t=1}\sum^Q_{q=1}I(\hat{\mathcal{Q}}_{\tau_q,t}\neq Sort(\hat{\mathcal{Q}}_{\tau_q,t}))$. Unfortunately, implementing this measure directly in the constraint would pose difficulties for any optimising algorithm on account of the measure being discrete. \citet{galvao2011threshold} proposes using an alternative way to measure quantile crossing:% follow \citet{galvao2011threshold} in focusing on the raw difference between every adjacent quantile:
\begin{equation} \label{eq:NCGalvao}
    Cross(\hat{\mathcal{Q}}_{\tau_q,t})=\min(0,\hat{\mathcal{Q}}_{\tau_q,t}-\hat{\mathcal{Q}}_{\tau_{q-1},t})
\end{equation}

The key to this measure is that it looks at the raw difference between adjacent quantiles. The series of quantiles $\tau$ is monotonically increasing by definition. As such, $\hat{\mathcal{Q}}_{\tau_q,t}-\hat{\mathcal{Q}}_{\tau_{q-1},t}$ is a negative value if and only if the quantiles cross at time $t$. Since, the measure is thresholded at 0, taking the sum over all $t$ gives the ``total distance of crossing'' for the two quantiles. This measure has several advantages. First, this measure of crossing a continuous and has the same units as the objective function in equation (\ref{eq:constrainedoptim}). As such can be included as the constraints without causing issues for an optimiser. Second, just like the objective function, this measure is easy to scale as the number of quantiles. increases: simply take the average (or sum) across all quantiles.

Third, equation (\ref{eq:NCGalvao}) is related to the constraints of the estimator described in \citet{bondell2010noncrossing}. To see this, %This equation provides a sufficient measure for crossing, but we wish to clearly establish the link to the estimated parameters. To this end 
define $\gamma_{\tau_q}=\beta_{\tau_q}-\beta_{\tau_{q-1}}$, i.e. the difference in the estimated parameters, and restrict $L=0$. In this case %If $L=0$, then 
$x_t^T\gamma_{\tau_q}$ is a sufficient measure of quantile differences. Next define $\gamma_{\tau_q}=\gamma_{\tau_q}^+-\gamma_{\tau_q}^-$, where $\gamma_{\tau_q}^+\geq0$ and $\gamma_{\tau_q}^-\geq0$. \citet{bondell2010noncrossing} propose only focusing on the worst case scenario so as to reduce the number of constraints from $T(Q-1)$ to $Q-1$. Following \citet{szendrei2023fused}, we can rewrite this worst case scenario as: $\max(X)\gamma_{\tau_q}^-+\min(X)\gamma_{\tau_q}^+$. If there is no quantile crossing, then this measure, just like the direct quantile difference in equation (\ref{eq:NCGalvao}), is positive. If quantile crossing occurs than both measures yield negative values. The key difference between the two measures is that \citet{galvao2011threshold} is specific to an observation, while \citet{bondell2010noncrossing} is not. As such, the two measures are comparable if we take the sum of equation (\ref{eq:NCGalvao}):

\begin{equation} \label{eq:BondellasPenalty}
    \sum^T_{t=1}\min(0,x_t^T[\gamma^+_{\tau_q}-\gamma^-_{\tau_q}]) \propto \min(0,\max(X)\gamma_{\tau_q}^-+\min(X)\gamma_{\tau_q}^+)
\end{equation}

Finally, equation (\ref{eq:NCGalvao}) can be used as a penalty in the penalty method of constraint optimisation \citep{boyd2004convex}. A function $P(u)$ is a valid penalty if it is (1) continuous; (2) $P(u)\geq0$ for all $u\in\mathbb{R}^n$, and; (3) $P(u)=0$ when the objective function is minimised within the constraint. Equation (\ref{eq:NCGalvao}) can be rewritten in such a way that it satisfies all three conditions:

\begin{equation} \label{eq:NCGalvaoReWrite}
    Cross(\hat{\mathcal{Q}}_{\tau_q,t})=\max(0,-[\hat{\mathcal{Q}}_{\tau_q,t}-\hat{\mathcal{Q}}_{\tau_{q-1},t}])
\end{equation}

The measure in equation (\ref{eq:NCGalvaoReWrite}) will be used as a measure of crossing at time $t$ between two adjacent quantiles. To make the connection between this measure and the estimated coefficients clear when $L\geq0$, we rewrite it as

\begin{equation} \label{eq:NC}
    Cross(\hat{\mathcal{Q}}_{\tau_q,t})=\max\Big(0,-\Big[x_t^T\gamma_{\tau_q}+\sum^L_{l=1}\Big[\hat{\mathcal{Q}}_{\tau_q,t-l}\theta_{\tau_q,l}-\hat{\mathcal{Q}}_{\tau_{q-1},t-l}\theta_{\tau_{q-1,l}}\Big]\Big]\Big)
\end{equation}

When $L=0$ equation (\ref{eq:NC}) becomes the left hand side of equation (\ref{eq:BondellasPenalty}). In this way equation (\ref{eq:NC}) is more general: it works for any degree of quantile inertia. The summation in this equation carries over part of the past quantile differences. In this way quantile inertia influences both the objective function and the constraint directly.

%non-crossing. However, when $L=1$, we need to introduce ``stickiness'' in the difference between adjacent quantiles. In particular, note that the degree of quantile difference rigidity is influenced by the quantile specific lagged quantile coeffficients. We define this quantile difference rigidity term as: $\Delta \hat{\mathcal{Q}}_{\tau_q,t-1}=\hat{\mathcal{Q}}_{\tau_q,t-1}\theta_{\tau_q}-\hat{\mathcal{Q}}_{\tau_{q-1},t-1}\theta_{\tau_{q-1}}$. Putting these terms together leads to the following measure of crossing:
%\begin{equation} \label{eq:NC}
%    Cross(\hat{\mathcal{Q}}_{\tau_q,t})=\min(0,[x_t^T\gamma_{\tau_q}+\Delta %\hat{\mathcal{Q}}_{\tau_q,t-1}]).
%\end{equation}

In this paper we will take the average of this measure over the estimated quantiles (minus 1) and the number of observations, $T$, to obtain an average crossing distance measure. Taking equation (\ref{eq:constrainedoptim}) and swapping the constraint to be the crossing measure in equation (\ref{eq:NC}) leads to the following function:

\begin{equation} \label{eq:constrainedoptimNC}
    \begin{split}
        \hat{\delta}&=\underset{\delta}{argmin}\frac{1}{QT}\sum^{Q}_{q=1}\sum^{T}_{t=1}\rho_{\tau_q}(y_{t}-x_t^T\beta_{\tau_q}-\hat{\mathcal{Q}}_{\tau_q,t-1}^T\theta_{\tau_q})\\
        s.t.~&\frac{1}{(Q-1)T}\sum^{Q}_{q=2}\sum^{T}_{t=1} \max\Big(0,-\Big[x_t^T\gamma_{\tau_q}+\sum^L_{l=1}\Big[\hat{\mathcal{Q}}_{\tau_q,t-l}\theta_{\tau_q,l}-\hat{\mathcal{Q}}_{\tau_{q-1},t-l}\theta_{\tau_{q-1,l}}\Big]\Big]\Big)
    \end{split}
\end{equation}

Now we only have a single constraint, which forces the number of crossings to be equal to 0. %We can switch the $\min$ in the constraint to a $\max$, which makes setting up optimisation easier, since now we need to minimise both the objective function and the crossing distance. We write the Lagrangian of the above equation to finally obtain our objective function:
As mentioned before, the measure of crossing distance is a valid penalty function and as such we can convert the constrained optimisation problem into a (penalised) unconstrained optimisation problem:

\begin{equation} \label{eq:DynQR}
\begin{split}
    \hat{\delta}&=\underset{\delta}{argmin}\frac{1}{QT}\sum^{Q}_{q=1}\sum^{T}_{t=1}\rho_{\tau_q}(y_{t}-x_t^T\beta_{\tau_q}-\hat{\mathcal{Q}}_{\tau_q,t-1}^T\theta_{\tau_q})\\
    +&\frac{\lambda}{(Q-1)T}\sum^{Q}_{q=2}\sum^{T}_{t=1} \max\Big(0,-\Big[x_t^T\gamma_{\tau_q}+\sum^L_{l=1}\Big[\hat{\mathcal{Q}}_{\tau_q,t-l}\theta_{\tau_q,l}-\hat{\mathcal{Q}}_{\tau_{q-1},t-l}\theta_{\tau_{q-1,l}}\Big]\Big]\Big)
\end{split}
\end{equation}

This equation is the crossing penalised multi-quantile CAViaR. The first part of the equation is the multi-equation CAViaR as in \citet{white2010modeling}, while the second part of the equation is the crossing distance penalty. On account of condensing the crossing distance into a single constraint, we end up with a scalar tuning parameter ($\lambda$). Using crossing distance in a penalised regression framework can help push the quantiles towards non-crossing as the tuning parameter is increased. In this way we end up with multi-equation CAViaR models that do not cross in-sample.

While obtaining the above objective function in equation (\ref{eq:DynQR}) was relatively straightforward, estimating it is much more difficult. The landscape of the objective function is quite ``rugged'' on account of the pinball loss function ($\rho_{\tau_q}(\cdot)$) as well as the crossing distance penalty. As such, it is anyway challenging to obtain a solution for cases when $\lambda=0$, but we render it even more difficult by imposing larger penalties for crossing. Furthermore, while it is possible to estimate the initial objective function in equation (\ref{eq:CAViaR}) via a Laplace type estimator (see \citet{chernozhukov2001conditional}, \citet{chen2012forecasting}, and \citet{rubia2013downside} among others), introduction of the crossing penalty makes this approach unavailable. In particular, while the residuals of equation (\ref{eq:CAViaR}) can be assumed to have a skewed Laplace distribution, which then allows the use of MCMC methods \citep{chernozhukov2003mcmc}, the same is not true for equation (\ref{eq:DynQR}). Hence, we use a blackbox optimisation routine called covariance matrix adaptation evolution strategy (hereafter CMA-ES) \citep{hansen2009tec}.

\subsection{Estimation}
In this section we provide a brief overview of the optimisation method used. While the aim here is to give an intuitive understanding of the algorithm, for further details, please refer to \citet{hansen2009tec} and \citet{audet2017derivative} and references therein.

The CMA-ES routine is an evolutionary algorithm commonly employed for difficult non-linear and non-convex optimisation problems without information on derivatives. Importantly, the routine works on the continuous domain, which partly motivates our choice of the measure of crossing in the objective function. Formulating the objective function as a penalised regression framework is also beneficial as the CMA-ES handles unconstrained or (parameter) bounded constraint optimisation problems well. 

The key advantages of the CMA-ES is that it can handle challenging optimisation scenarios where the problem is non-separable, meaning that variables are highly dependent on each other, or badly conditioned, where the curvature of the objective function varies significantly \citep{hansen2001completely}. Furthermore, because the routine does not use gradient information, it is a feasible method for non-smooth and non-continuous objective functions. This characteristic is particularly valuable when dealing with noisy objective functions or problems with multiple local optima \citep{hansen2009benchmarking}. In short, the CMA-ES is best suited if derivative based methods fail due to a ``rugged'' search landscape, which is likely the case for our proposed objective function here. Furthermore, the inclusion of the crossing penalty makes the objective function non-separable.\footnote{If we had allowed the quantiles to influence each other, then the joint estimation framework would be non-separable even when $\lambda=0$ in equation (\ref{eq:DynQR}).}

CMA-ES is a second order optimisation method, estimating a positive definite matrix in an iterative procedure. Unlike quasi-Newton methods that rely on gradient information, the CMA-ES estimates the covariance matrix, which is related to the inverse Hessian, to guide the search. An evolutionary heuristic is used to discard the worst sampled points, and only calculate the covariance matrix based on the best fitted points. The method has been shown to not just be an adequate local optimisation routine but also a global optimisation method \citep{hansen2004evaluating}.

Additionally, the CMA-ES exhibits key invariance properties. It remains unchanged when the objective function undergoes order-preserving transformations, i.e., strictly monotonic changes, or when the search space undergoes angle-preserving transformations like rotation, reflection, or translation. These invariances ensure that the algorithm behaves consistently across different classes of functions.

The key to the CMA-ES algorithm is that it samples from a multivariate normal distribution. This multivariate normal gets updated with each iteration, and the sampled points get closer to the solution with each iteration. %\footnote{Following \citet{engle2004caviar}, \citet{white2010modeling}, and \citet{white2015var}, the multivariate normal is a good candidate distribution to sample from because it induces asymptotic normality of the CAViaR parameters.} 
To start the procedure, one needs to provide an initial mean vector ($\mu$) and covariance matrix ($C$), which determines the initial search space. With these pre-specified parameters a population ($p_i$) is generated randomly from a multivariate normal $\mathcal{N}(\mu,C)$. Because the function evaluation is relatively cheap computationally and can be parallelised, we set up a relatively large population size of $\max(100,K \times 10)$ where $K$ is the number of parameters being estimated (including intercepts). Note that due to joint estimation, $K$ can be relatively large even when the number of variables is limited but the number of quantiles being estimated is large. All sampled points are evaluated using the objective function of equation (\ref{eq:DynQR}).

The evolutionary heuristic enters the algorithm at this point: We select the top quarter of sampled points to update $\mu$ and $C$. To update the mean vector, $\mu$, we use the following formula:

\begin{equation}
    \mu_{new}=\sum^{\frac{\max(100,K \times 10)}{4}}_{i=1} w_i p_i,
\end{equation}
where $w_i$ is a weight assigned to the ``offspring''. The weight is based on the rank of the sampled point, and it decreases superlinearly. A new covariance matrix $C_{new}$ is then obtained using $\mu_{new}$. Note that just like in the case for the mean vector, the covariance matrix update also factors in the rank of the ``offspring'' via the weight function. With the parameters updated a new sample can be generated using:

\begin{equation} \label{eq:CMA-ESearch}
    p_{i,new}=\mu_{new}+\sigma\mathcal{N}(0,C_{new}).
\end{equation}

The value of $\sigma$ controls the magnitude of the step taken in the update and is scaled by the success of the current step \citep{hansen2006cma}.\footnote{Here, we provide only a general overview of the algorithm and do not discuss iterative updates for $C$ and updates for $\sigma$. For further information on these steps and the algorithm please see \citet{hansen2016cma} and references therein.} Next the algorithm repeats these steps until convergence is achieved or a preset value of maximum iterations is reached. Equation (\ref{eq:CMA-ESearch}) highlights how the CMA-ES is a stochastic search algorithm guided by an elitist evolutionary heuristic. 

\newpage
\section{Monte Carlo}
\subsection{Experimental design}
To construct the data for our Monte Carlo experiments, we follow the method outlined in \citet{white2010modeling}. To this end, we need to define a vector for the parameters of interest: $\vartheta=\{\beta,\theta\}$. Since the parameters can vary by quantile, we also define a vector $\xi$, which will regulate how much the specific covariate varies with quantiles. In essence the coefficient's value given a specific quantile can be defined following \citet{koenker2006quantile}:

\begin{equation} \label{eq:MCparam}
    \vartheta(\tau)=\vartheta+\xi F_\varepsilon^{-1}(\tau),
\end{equation}
where $F_\varepsilon$ is the cumulative density function of the error, which we assume to be Gaussian (normally distributed), and $\tau$ represents the quantile which is follows the uniform distribution between 0 and 1. For the simulations we fix $\vartheta=\{2,0.5,-3,0.25\}$, where the first entry corresponds to the intercept, the second entry is the effect of lag of the variable (whose quantiles are being estimated), the third entry the impact of an exogenous variable (which follows a uniform distribution between 0 and 1), and the final entry corresponds to coefficient of the lag of the quantile itself. The quantile variation for each of these variables is fixed at $\xi=\{1,0.15,1,\frac{0.15}{2}\}$. Note that the lagged variable and lagged quantiles quantile variation is relatively small. This is set to ensure that the simulated upper quantiles remain stationary.

There are three designs created with varying degree of heteroskedasticity. $y_1$ is a homoskedastic design where all entries of $\xi$ are 0 except the first, which relates to the intercept. In $y_2$, the exogenous variable is allowed to vary along with the intercept, i.e. all entries of $\xi$ are 0 except the first and third. Finally, in $y_3$ all coefficients are quantile varying. Furthermore, we test two types of processes: QAR(1) and CAViaR(1). In the QAR(1) case, the lagged quantile coefficient is constrained to 0 for all quantiles, while for CAViaR(1) the lagged quantile is included in the DGP. Note that this means that the QAR(1) will have $\tau$ fewer parameters if the models are correctly specified during the estimation. 

%\newpage
In total, we simulate 50 Monte Carlo datasets generated with two sample sizes $T=\{50,200\}$. Since we also estimate the CAViaR model as described in \citet{white2010modeling} and \cite{engle2004caviar}, we refer to our CAViaR \textit{estimator} as CAViaR-NM, and denote by DQAR(1,1) the existing CAViaR \textit{process}. Here, DQAR($m$,$l$) stands for Dynamic Quantile Autoregression where $m$ is the number of lags of the dependent variable and $l$ is the number of lags of the quantile.

Given the coefficient matrix for each quantile, data were generated following the method described in \citet{white2010modeling}. Given the values of the quantiles, the observation, and the exogenous variable in $t-1$, we compute each $10^{th}$ percentile (quantiles $q=0.001, 0.002, \ldots, 0.999$). We then draw a random variable, $U_t$, from a uniform distribution between 0 and 1. Given $U_t$, we evaluate where in the distribution this draw falls given our estimated conditional quantiles. Next we draw from a uniform distribution within the interval given by our quantiles. Formally, this procedure is:

\begin{equation}
    y_t=\sum^{Q+1}_{q=1}\mathcal{I}(u_{q-1}<U_t<u_q)[\hat{\mathcal{Q}}_{\tau_q,t}+V_t(\hat{\mathcal{Q}}_{\tau_q,t}-\hat{\mathcal{Q}}_{\tau_{q-1},t})],
\end{equation}
where $U_t\sim U(0,1)$, $V_t\sim U(0,1)$, and $\hat{\mathcal{Q}}_{\tau_q,t}$ is calculated following equation (\ref{eq:CAViaR}). When we draw from the extreme quantiles, i.e. draws below the $0.1^{th}$ quantile ($q=0$) or above the $99.9^{th}$ quantile ($q=Q+1$), we generate the boundary by taking the lowest (or highest) quantile and substracting (or adding) $0.001^{th}$ of the inter quartile range at time $t$. The process iterates until $t=T+50$. We set all initial values (lagged quantiles and lags of variable) to be 0. Due to the initial point potentially influencing the first few observations, we discard the first 50 simulated data points. To ensure that there is no crossing in the ideal quantiles, we check the simulated series for crossing and re-simulate with new randomly generated variables if there is any crossing. Note that it does not matter that the $U_t$ and $V_t$ are drawn from a uniform distribution, since the parameters are constructed via equation (\ref{eq:MCparam}), and as such the quantile profile is dictated by $F_\varepsilon^{-1}(\tau)$. The reason we need this somewhat convoluted Monte Carlo setup is because of the lagged quantiles influencing the dependent variable.

While we simulate over 9000 quantiles, we only estimate 9 quantiles, i.e. every $10^{th}$ quantile, to evaluate the performance of the different estimators. Specifically, we compare the traditional QR of \citet{koenker1978regression}, the non-crossing quantile regression of \citet{bondell2010noncrossing}, the multiple quantile CaViaR estimator of \citet{white2010modeling}, and the proposed crossing penalised CAViaR method. We use average coefficient bias as the central measure of performance for the different estimators. We also consider the average number of crossing incidences for the different estimators. This measure is computed by comparing unsorted and sorted fitted quantiles and counting the instances the two do not match. Taking the overall average of this crossing incidence provides a measure of the probability of crossing. Ideally, a good estimator should have low values on both measures. 

Furthermore, because the initial point for derivative-free optimisation routines can have undue influence on the solution, we estimate and report the coefficient bias and crossing incidence for the crossing penalised CAViaR model with 3 different initial conditions. We also estimate the canonical CAViaR model (without crossing penalisation), calculated as described in \citet{white2010modeling}, with the same initial conditions.\footnote{To ensure that the lagged quantile coefficients do not go above 1, we use the NM implementation of \citet{fminsearchbnd}, where one can impose boundary constraints for the parameters.}$^{,}$\footnote{The CAViaR model described in \citet{engle2004caviar} first creates a grid of initial conditions and selects the best $n$ candidates to run the NM routine. Here we set the QR and BRW coefficients as given, and create a grid of 1000 potential values for the lagged quantile coefficients drawn from a uniform distribution. Then, we select the best candidate (i.e. $n=1$), to run the NM algorithm on.}

In total we present 9 types of DynQR (crossing penalised CAViaR with CMA-ES) estimates (3 different $\lambda$ values, each with 3 different initial conditions), 3 types of CAViaR estimates (3 different initial conditions, denoted as CAViaR-NM), and 2 regular quantile regression methods: QR of \citet{koenker1978regression}; and non-crossing QR of \citet{bondell2010noncrossing}, denoted as BRW. The results for the coefficient bias of the QAR(1) model are presented in table (\ref{tab:QARCoeffBias}), while the coefficient bias of the DAQR(1,1) DGP are shown in table (\ref{tab:CAViaRCoeffBias}). The probability of crossing is summarised in table (\ref{tab:Crossnum}) for both types of processes.

\subsection{Results}

Evaluating the QAR(1) coefficient bias results in table (\ref{tab:QARCoeffBias}) reveals a few interesting points. First, the DynQR with $\lambda=0$ yields the same coefficient bias as the regular QR. This is reassuring and highlights that the CMA-ES routine can be used to recover the QR coefficients. Furthermore, DynQR is not impacted by the initial conditions provided, as it leads to the same coefficient bias. Importantly, these observations are true for both sample sizes. However, the choice of penalty parameter has a large influence for the coefficient bias. As the hyperparameter is increased the coefficient bias decreases, with $\lambda=5$ yielding very close coefficient bias to the BRW. Note that this finding is largely on account of correctly specifying the model. If the model were misspecified, enforcing non-crossing would not necessarily lead to lower coefficient bias. In particular, due to non-crossing constraints being a type of fused-shrinkage, imposing non-crossing constrainst on a misspecified model might lead to overshrinking quantile variation as outlined in \citet{szendrei2023fused}.

\begin{landscape}
\begin{table}[]
\centering
\resizebox{1.3\textwidth}{!}{%
\begin{tabular}{ll|ccc|ccc|ccc|ccc|cc}
\hline
 &  & \multicolumn{3}{c|}{DynQR $(\lambda=0)$} & \multicolumn{3}{c|}{DynQR $(\lambda=1)$} & \multicolumn{3}{c|}{DynQR $(\lambda=5)$} & \multicolumn{3}{c|}{CAViaR-NM} & \multicolumn{2}{c}{Regular QR} \\
 &  & $p_0=$0 & $p_0=$QR & $p_0=$BRW & $p_0=$0 & $p_0=$QR & $p_0=$BRW & $p_0=$0 & $p_0=$QR & $p_0=$BRW & $p_0=$0 & $p_0=$QR & $p_0=$BRW & BRW & QR \\ \hline \hline
\multicolumn{2}{l|}{\textbf{T=50}} &  &  &  &  &  &  &  &  &  &  &  &  &  &  \\
 & $y_1(\tau=0.1)$ & 0.490 & 0.490 & 0.490 & 0.465 & 0.465 & 0.465 & 0.463 & 0.463 & 0.463 & 0.858 & 0.490 & 0.486 & 0.461 & 0.490 \\
 & $y_1(\tau=0.5)$ & 0.376 & 0.376 & 0.376 & 0.359 & 0.359 & 0.359 & 0.359 & 0.359 & 0.359 & 0.524 & 0.376 & 0.376 & 0.355 & 0.376 \\
 & $y_1(\tau=0.9)$ & 0.574 & 0.574 & 0.574 & 0.530 & 0.530 & 0.530 & 0.530 & 0.530 & 0.530 & 0.701 & 0.574 & 0.574 & 0.509 & 0.574 \\
 &  &  &  &  &  &  &  &  &  &  &  &  &  &  &  \\
 & $y_2(\tau=0.1)$ & 0.680 & 0.680 & 0.680 & 0.603 & 0.603 & 0.603 & 0.602 & 0.602 & 0.602 & 1.031 & 0.680 & 0.681 & 0.595 & 0.680 \\
 & $y_2(\tau=0.5)$ & 0.456 & 0.456 & 0.456 & 0.454 & 0.454 & 0.454 & 0.454 & 0.454 & 0.454 & 0.537 & 0.456 & 0.458 & 0.433 & 0.456 \\
 & $y_2(\tau=0.9)$ & 0.706 & 0.706 & 0.706 & 0.689 & 0.689 & 0.689 & 0.689 & 0.689 & 0.689 & 0.804 & 0.706 & 0.706 & 0.652 & 0.706 \\
 &  &  &  &  &  &  &  &  &  &  &  &  &  &  &  \\
 & $y_3(\tau=0.1)$ & 0.942 & 0.942 & 0.942 & 0.869 & 0.869 & 0.869 & 0.869 & 0.869 & 0.869 & 1.213 & 0.942 & 0.939 & 0.829 & 0.942 \\
 & $y_3(\tau=0.5)$ & 0.539 & 0.539 & 0.539 & 0.522 & 0.522 & 0.522 & 0.522 & 0.522 & 0.522 & 0.703 & 0.539 & 0.537 & 0.522 & 0.539 \\
 & $y_3(\tau=0.9)$ & 0.697 & 0.697 & 0.697 & 0.694 & 0.694 & 0.694 & 0.694 & 0.694 & 0.694 & 0.790 & 0.697 & 0.708 & 0.676 & 0.697 \\ \hline
\multicolumn{2}{l|}{\textbf{T=200}} &  &  &  &  &  &  &  &  &  &  &  &  &  &  \\
 & $y_1(\tau=0.1)$ & 0.236 & 0.236 & 0.236 & 0.235 & 0.235 & 0.235 & 0.235 & 0.235 & 0.235 & 0.357 & 0.236 & 0.235 & 0.232 & 0.236 \\
 & $y_1(\tau=0.5)$ & 0.189 & 0.189 & 0.189 & 0.187 & 0.187 & 0.187 & 0.187 & 0.187 & 0.187 & 0.187 & 0.189 & 0.188 & 0.184 & 0.189 \\
 & $y_1(\tau=0.9)$ & 0.252 & 0.252 & 0.252 & 0.250 & 0.250 & 0.250 & 0.250 & 0.250 & 0.250 & 0.254 & 0.252 & 0.252 & 0.246 & 0.252 \\
 &  &  &  &  &  &  &  &  &  &  &  &  &  &  &  \\
 & $y_2(\tau=0.1)$ & 0.324 & 0.324 & 0.324 & 0.326 & 0.326 & 0.326 & 0.326 & 0.326 & 0.326 & 0.492 & 0.324 & 0.324 & 0.318 & 0.324 \\
 & $y_2(\tau=0.5)$ & 0.253 & 0.253 & 0.253 & 0.253 & 0.253 & 0.253 & 0.253 & 0.253 & 0.253 & 0.251 & 0.253 & 0.253 & 0.247 & 0.253 \\
 & $y_2(\tau=0.9)$ & 0.301 & 0.301 & 0.301 & 0.297 & 0.298 & 0.297 & 0.298 & 0.298 & 0.297 & 0.384 & 0.301 & 0.301 & 0.295 & 0.301 \\
 &  &  &  &  &  &  &  &  &  &  &  &  &  &  &  \\
 & $y_3(\tau=0.1)$ & 0.321 & 0.321 & 0.321 & 0.319 & 0.319 & 0.319 & 0.319 & 0.319 & 0.319 & 0.467 & 0.321 & 0.323 & 0.315 & 0.321 \\
 & $y_3(\tau=0.5)$ & 0.196 & 0.196 & 0.196 & 0.190 & 0.190 & 0.190 & 0.187 & 0.187 & 0.187 & 0.229 & 0.196 & 0.195 & 0.188 & 0.196 \\
 & $y_3(\tau=0.9)$ & 0.351 & 0.351 & 0.351 & 0.347 & 0.347 & 0.347 & 0.347 & 0.347 & 0.347 & 0.355 & 0.351 & 0.351 & 0.346 & 0.351 \\ \hline
\end{tabular}%
}
\caption{Coefficient bias for QAR(1)}
\label{tab:QARCoeffBias}
\end{table}
\end{landscape}

\begin{landscape}
\begin{table}[]
\centering
\resizebox{1.3\textwidth}{!}{%
\begin{tabular}{ll|ccc|ccc|ccc|ccc|cc}
\hline
 &  & \multicolumn{3}{c|}{DynQR $(\lambda=0)$} & \multicolumn{3}{c|}{DynQR $(\lambda=1)$} & \multicolumn{3}{c|}{DynQR $(\lambda=5)$} & \multicolumn{3}{c|}{CAViaR-NM} & \multicolumn{2}{c}{Regular QR} \\
 &  & $p_0=$0 & $p_0=$QR & $p_0=$BRW & $p_0=$0 & $p_0=$QR & $p_0=$BRW & $p_0=$0 & $p_0=$QR & $p_0=$BRW & $p_0=$0 & $p_0=$QR & $p_0=$BRW & BRW & QR \\ \hline \hline
\multicolumn{2}{l|}{\textbf{T=50}} &  &  &  &  &  &  &  &  &  &  &  &  &  &  \\
 & $y_1(\tau=0.1)$ & 0.599 & 0.599 & 0.599 & 0.585 & 0.585 & 0.585 & 0.583 & 0.583 & 0.583 & 1.101 & 0.557 & 0.565 & 0.533 & 0.568 \\
 & $y_1(\tau=0.5)$ & 0.447 & 0.444 & 0.446 & 0.424 & 0.424 & 0.424 & 0.424 & 0.424 & 0.431 & 1.124 & 0.447 & 0.461 & 0.455 & 0.466 \\
 & $y_1(\tau=0.9)$ & 0.612 & 0.623 & 0.623 & 0.594 & 0.594 & 0.583 & 0.580 & 0.595 & 0.583 & 1.707 & 0.690 & 0.680 & 0.678 & 0.713 \\
 &  &  &  &  &  &  &  &  &  &  &  &  &  &  &  \\
 & $y_2(\tau=0.1)$ & 0.913 & 0.980 & 0.898 & 0.833 & 0.892 & 0.896 & 0.833 & 0.847 & 0.848 & 1.433 & 0.977 & 1.014 & 0.937 & 0.986 \\
 & $y_2(\tau=0.5)$ & 0.557 & 0.573 & 0.574 & 0.602 & 0.604 & 0.605 & 0.599 & 0.618 & 0.621 & 1.116 & 0.582 & 0.581 & 0.628 & 0.651 \\
 & $y_2(\tau=0.9)$ & 0.884 & 0.870 & 0.907 & 0.889 & 0.916 & 0.911 & 0.880 & 0.896 & 0.911 & 1.435 & 0.978 & 1.007 & 0.966 & 0.980 \\
 &  &  &  &  &  &  &  &  &  &  &  &  &  &  &  \\
 & $y_3(\tau=0.1)$ & 0.971 & 0.961 & 0.971 & 0.968 & 0.951 & 0.956 & 0.956 & 0.946 & 0.955 & 1.477 & 1.040 & 1.020 & 1.016 & 1.099 \\
 & $y_3(\tau=0.5)$ & 0.887 & 0.859 & 0.854 & 0.852 & 0.819 & 0.848 & 0.849 & 0.844 & 0.815 & 1.457 & 0.878 & 0.884 & 0.874 & 0.957 \\
 & $y_3(\tau=0.9)$ & 1.512 & 1.591 & 1.591 & 1.403 & 1.428 & 1.430 & 1.427 & 1.428 & 1.426 & 1.620 & 1.791 & 1.773 & 1.822 & 1.915 \\ \hline
\multicolumn{2}{l|}{\textbf{T=200}} &  &  &  &  &  &  &  &  &  &  &  &  &  &  \\
 & $y_1(\tau=0.1)$ & 0.311 & 0.311 & 0.311 & 0.309 & 0.309 & 0.309 & 0.309 & 0.309 & 0.309 & 0.700 & 0.331 & 0.330 & 0.353 & 0.352 \\
 & $y_1(\tau=0.5)$ & 0.199 & 0.199 & 0.198 & 0.198 & 0.198 & 0.198 & 0.199 & 0.198 & 0.199 & 0.595 & 0.199 & 0.198 & 0.277 & 0.280 \\
 & $y_1(\tau=0.9)$ & 0.335 & 0.334 & 0.333 & 0.323 & 0.323 & 0.323 & 0.323 & 0.323 & 0.323 & 1.294 & 0.348 & 0.350 & 0.428 & 0.433 \\
 &  &  &  &  &  &  &  &  &  &  &  &  &  &  &  \\
 & $y_2(\tau=0.1)$ & 0.446 & 0.446 & 0.447 & 0.445 & 0.445 & 0.445 & 0.445 & 0.446 & 0.445 & 0.901 & 0.483 & 0.488 & 0.544 & 0.544 \\
 & $y_2(\tau=0.5)$ & 0.304 & 0.304 & 0.304 & 0.301 & 0.302 & 0.303 & 0.302 & 0.302 & 0.301 & 0.813 & 0.308 & 0.306 & 0.343 & 0.340 \\
 & $y_2(\tau=0.9)$ & 0.476 & 0.484 & 0.482 & 0.470 & 0.469 & 0.469 & 0.469 & 0.470 & 0.470 & 1.114 & 0.521 & 0.521 & 0.592 & 0.593 \\
 &  &  &  &  &  &  &  &  &  &  &  &  &  &  &  \\
 & $y_3(\tau=0.1)$ & 0.422 & 0.423 & 0.423 & 0.394 & 0.394 & 0.394 & 0.390 & 0.391 & 0.391 & 0.964 & 0.410 & 0.427 & 0.398 & 0.367 \\
 & $y_3(\tau=0.5)$ & 0.338 & 0.338 & 0.338 & 0.327 & 0.327 & 0.327 & 0.330 & 0.330 & 0.330 & 0.946 & 0.337 & 0.333 & 0.417 & 0.433 \\
 & $y_3(\tau=0.9)$ & 0.673 & 0.686 & 0.686 & 0.681 & 0.689 & 0.685 & 0.680 & 0.684 & 0.682 & 1.280 & 0.725 & 0.718 & 1.246 & 1.243 \\ \hline
\end{tabular}%
}
\caption{Coefficient bias for DQAR(1,1)}
\label{tab:CAViaRCoeffBias}
\end{table}
\end{landscape}

Comparing CAViaR-NM with DynQR (with $\lambda=0$) shows a stark difference in performance. CAViaR-NM is influenced by the initial conditions. In fact when the inital conditions are a vector of 0's, the CAViaR-NM is beaten by conventional quantile estimators QR and BRW. This stark contrast highlights the global optimisation power of the CMA-ES, which is useful in many applications: it is difficult to know ex ante what initial conditions are optimal for the application at hand.

Considering the crossing probabilities for the QAR(1) case, shown on the left side of table (\ref{tab:Crossnum}), emphasises a few further observations. In particular, CAViaR with initial conditions as a vector of 0's gives the highest probability of crossing. However, when we provide CAViaR-NM the QR coefficients as initial conditions, the routine yields marginal gains in coefficient bias without making the probability of crossing worse. Interestingly, when giving the routine BRW coefficients as initial conditions, the routine yields worse crossing and coefficient bias results than the BRW. This is expected from the discussion in \citet{szendrei2023fused} that conceptualises non-crossing constraint tightness through the lens of bias-variance trade-off. Specifically, without a penalised regression framework, the objective function in equation (\ref{eq:CAViaR}) will tend to yield the solution that is $\lambda=0$, aiming to obtain the best in-sample fit.

Looking at crossing probability and coefficient bias of the DynQR, we can see that for most cases $\lambda=1$ is sufficient to yield low crossing probability and coefficient bias. This is true for all initial conditions given, and both sample sizes considered. Setting $\lambda=5$ results in near zero crossing probabilities, but with only marginal gains in coefficient bias. If one is confident that their model is not misspecified, then these marginal gains might be worthwhile, especially if sample size is small.

Interestingly, DynQR with $\lambda=0$ has a lower crossing probability than the traditional QR, while having the exact same coefficient bias as QR. Nevertheless, the differences in crossing probabilities between the QR and DynQR with $\lambda=0$, is less than 0.5\%. As such, we can still be assured that the CMA-ES routine is capable of recovering the traditional QR estimates in a QAR($p$) setting. 

Moving on to the DQAR(1,1) case we can see that the results from the QAR(1) case largely follow. Although the initial condition has a larger influence on the DynQR than in the QAR(1) case, especially in the small sample sizes, the impact of the hyperparameter is still larger. Furthermore, the gains in coefficient bias are not guaranteed when $\lambda$ is increased. In particular, the coefficient bias at the median increases for $y_3$ for both sample sizes when QR is given as an initial condition. However, increasing the penalty still yields better coefficient bias at the tails. This further corroborates the conclusion from the QAR(1) setting, that for most applications setting $\lambda=1$ is sufficient.

\begin{table}[t]
\centering
\resizebox{\textwidth}{!}{%
\begin{tabular}{ll|cccccc|cccccc}
\hline
 &  & \multicolumn{6}{c|}{QAR(1)} & \multicolumn{6}{c}{DQAR(1,1)} \\
 &  & \multicolumn{3}{c|}{T=50} & \multicolumn{3}{c|}{T=200} & \multicolumn{3}{c|}{T=50} & \multicolumn{3}{c}{T=200} \\
  &  & $y_1$ & $y_2$ & \multicolumn{1}{c|}{$y_3$} & $y_1$ & $y_2$ & $y_3$ & $y_1$ & $y_2$ & \multicolumn{1}{c|}{$y_3$} & $y_1$ & $y_2$ & $y_3$ \\ \hline \hline
\multicolumn{2}{l|}{DynQR ($\lambda=0$)} &  &  & \multicolumn{1}{c|}{} &  &  &  &  &  & \multicolumn{1}{c|}{} &  &  &  \\
 & $p_0=$0 & 7.022 & 7.747 & \multicolumn{1}{c|}{7.484} & 0.483 & 0.594 & 0.796 & 10.858 & 11.822 & \multicolumn{1}{c|}{11.124} & 0.922 & 1.353 & 1.932 \\
 & $p_0=$QR & 7.000 & 7.769 & \multicolumn{1}{c|}{7.520} & 0.487 & 0.596 & 0.804 & 10.849 & 11.787 & \multicolumn{1}{c|}{11.498} & 0.911 & 1.329 & 1.912 \\
 & $p_0=$BRW & 6.973 & 7.751 & \multicolumn{1}{c|}{7.529} & 0.484 & 0.594 & 0.797 & 10.800 & 11.756 & \multicolumn{1}{c|}{11.804} & 0.920 & 1.344 & 1.927 \\
\multicolumn{2}{l|}{DynQR ($\lambda=1$)} &  &  & \multicolumn{1}{c|}{} &  &  &  &  &  & \multicolumn{1}{c|}{} &  &  &  \\
 & $p_0=$0 & 0.182 & 0.258 & \multicolumn{1}{c|}{0.382} & 0.030 & 0.024 & 0.111 & 0.676 & 0.724 & \multicolumn{1}{c|}{0.831} & 0.050 & 0.100 & 0.230 \\
 & $p_0=$QR & 0.178 & 0.311 & \multicolumn{1}{c|}{0.404} & 0.033 & 0.024 & 0.117 & 0.604 & 0.742 & \multicolumn{1}{c|}{0.818} & 0.040 & 0.091 & 0.228 \\
 & $p_0=$BRW & 0.160 & 0.324 & \multicolumn{1}{c|}{0.391} & 0.031 & 0.024 & 0.114 & 0.613 & 0.711 & \multicolumn{1}{c|}{0.849} & 0.050 & 0.090 & 0.218 \\
\multicolumn{2}{l|}{DynQR ($\lambda=5$)} &  &  & \multicolumn{1}{c|}{} &  &  &  &  &  & \multicolumn{1}{c|}{} &  &  &  \\
 & $p_0=$0 & 0.018 & 0.009 & \multicolumn{1}{c|}{0.027} & 0.000 & 0.000 & 0.000 & 0.053 & 0.098 & \multicolumn{1}{c|}{0.129} & 0.002 & 0.000 & 0.010 \\
 & $p_0=$QR & 0.053 & 0.013 & \multicolumn{1}{c|}{0.009} & 0.000 & 0.000 & 0.002 & 0.049 & 0.018 & \multicolumn{1}{c|}{0.124} & 0.000 & 0.000 & 0.009 \\
 & $p_0=$BRW & 0.000 & 0.000 & \multicolumn{1}{c|}{0.000} & 0.000 & 0.000 & 0.000 & 0.058 & 0.031 & \multicolumn{1}{c|}{0.111} & 0.002 & 0.002 & 0.007 \\
\multicolumn{2}{l|}{CAViaR-NM} &  &  & \multicolumn{1}{c|}{} &  &  &  &  &  & \multicolumn{1}{c|}{} &  &  &  \\
 & $p_0=$0 & 18.609 & 18.182 & \multicolumn{1}{c|}{17.022} & 4.018 & 3.023 & 5.477 & 44.782 & 36.951 & \multicolumn{1}{c|}{36.400} & 32.680 & 29.931 & 26.559 \\
 & $p_0=$QR & 7.453 & 8.298 & \multicolumn{1}{c|}{8.080} & 0.493 & 0.603 & 0.813 & 11.138 & 12.764 & \multicolumn{1}{c|}{11.244} & 1.024 & 1.394 & 1.916 \\
 & $p_0=$BRW & 6.613 & 7.591 & \multicolumn{1}{c|}{7.409} & 0.477 & 0.578 & 0.784 & 10.769 & 12.031 & \multicolumn{1}{c|}{11.187} & 1.131 & 1.361 & 1.929 \\
\multicolumn{2}{l|}{Regular QR} &  &  & \multicolumn{1}{c|}{} &  &  &  &  &  & \multicolumn{1}{c|}{} &  &  &  \\
 & BRW & 0.000 & 0.000 & \multicolumn{1}{c|}{0.000} & 0.000 & 0.000 & 0.002 & 0.000 & 0.000 & \multicolumn{1}{c|}{0.009} & 0.000 & 0.000 & 0.000 \\
 & QR & 7.453 & 8.298 & \multicolumn{1}{c|}{8.080} & 0.493 & 0.603 & 0.813 & 7.511 & 7.698 & \multicolumn{1}{c|}{7.311} & 0.336 & 0.458 & 1.009 \\ \hline
\end{tabular}%
}
\caption{Probability of crossing}
\label{tab:Crossnum}
\end{table}

The influence of initial condition on the CAViaR-NM is even greater for the DQAR(1,1) setting than in the QAR(1) setting. In particular, giving the method the initial conditions of a vector of 0's results in such poor coefficient bias results that it only beats the QR and BRW, which are methods that do not account for lagged quantile dynamics. Furthermore, DynQR with $\lambda>0$ often provides better coefficient bias in the DQAR(1,1) setting with $T=50$ than CAViaR-NM with $T=200$ when initial conditions is a vector of 0's.

%\newpage
All estimators result in better performance across both metrics when more data are available. Nevertheless, the degree of improvement is not uniform across the different DGP's. In particular, the more heteroskedastic the design, the greater the advantages of additional data. Furthermore, in the DQAR(1,1) case the improvement is not as great for the regular QR setups, which is understandable given that we would then be estimating misspecified models with these estimators.

Note also that the BRW yields non-crossing quantiles in the DQAR(1,1) setting, but its coefficient bias results are often beaten by both the DynQR (for all $\lambda$ values) and CAViaR-NM (with QR or BRW as initial conditions). Furthermore, we can see that even the simple QR yields better coefficient bias for $y_3$ when $T=200$. This emphasises that non-crossing quantiles should only be enforced when we are confident in our model specification. This gives credence to thinking about non-crossing in a penalised regression framework: we only wish to penalise crossing in so far as it yields better out of sample performance, as motivated in \citet{szendrei2023fused}.

While the superior performance of the DynQR is clear from these tables, the improvements over the CAViaR-NM come at a cost. In particular, the computation time for the DynQR is much longer than alternative estimators.\footnote{We conducted all computations on a single core of an Intel i5-7500 CPU at 3.40GHz.} The regular QR estimators need less than a second to find a solution, while the CAViaR-NM requires about 10-20 seconds (depending on sample size). However, the DynQR with $\lambda=0$ needs 40 seconds for the small sample size, and 120 seconds for the large sample size. When lagged quantile dynamics are allowed for, the DynQR needs up to 220 seconds on average. As mentioned in the previous section, the landscape becomes more ``rugged'' as the crossing penalty is increased, which in turn increases the time the optimisation routine needs to find a solution. When $\lambda=5$, the routine needs on average 50 seconds for the small sample size with no lagged quantile dynamics, and 240 seconds for the large sample size with lagged quantile dynamics. 

%We conducted all computations on a single core of an Intel i5-7500 CPU at 3.40GHz. 

While the time required by the DynQR seems somewhat daunting, we note that these computation times can be reduced significantly by parallelisation. Due to this, a practitioner with similar problems will likely only need a fraction of the time reported here to find estimates using the DynQR. Nevertheless, even with parallelisation, the computing time required to find a final solution is likely to still be higher than the regular QR estimators.

\section{Value-at-Risk application}

The series of daily market returns of the FTSE100 index is computed as the difference of log market index values on consecutive trading days. The sample has $T=254$ observations between January, 2, 2008, and December, 31, 2008. The year of 2008 was chosen on account of the Global Financial Crisis leading to higher volatilities on the markets at the end of the sample. The change in volatility allows us to gauge the performance of different quantile estimators in a dynamically changing market environment.

We will estimate the asymmetric slope specification of \citet{engle2004caviar}:

\begin{equation}
    \hat{\mathcal{Q}}_{\tau_q,t}=\alpha_{\tau_q}+|y_{t-1}^T|^{+}\beta^{+}_{\tau_q}+|y_{t-1}^T|^{-}\beta^{-}_{\tau_q}+\hat{\mathcal{Q}}_{\tau_q,t-1}^T\theta_{\tau_q}+\varepsilon_t.
\end{equation}

\noindent where $\alpha$ is the constant, $|y_{t-1}|^{i}$ is the absolute value of past return if the sign is $i$ and 0 otherwise. Without the ``sticky'' quantile captured by $\mathcal{Q}_{\tau_q,t-l}^T\theta_{\tau_q,l}$, the above equation is simply a type of QAR(1) model with asymmetric slope depending on the sign of past observations. The key idea behind including $\mathcal{Q}_{\tau_q,t}^T\theta_{\tau_q}$ in the estimation is that distribution for $t+h$ is influenced by the GDP distribution at $t$. Throughout this section we will focus on $h=1$, i.e. one (trading) day ahead forecasts.

In addition to in-sample fits we also examine out-of-sample performance of the estimators. These forecasts are computed on an expanding window basis where the initial in-sample period uses the first 100 observations of the sample, which constitutes $(254-100-h)$ forecast windows.

For the FTSE Valut-at-Risk with lagged quantiles, our central estimates are based on DynQR, with the 3 $\lambda$ values considered in the Monte Carlo study. Since initial conditions had limited impact upon the DynQR results, we only consider the case where the algorithm starts from the zero vector. We also estimate the CAViaR-NM, starting from an initial condition of QR coefficients. Here, we use QR coefficients as a starting point since in our Monte Carlo experiments, this often led to the best or close to the best CAViaR-NM results. Of the regular quantile estimators, we also present the results of the traditional QR, and the non-crossing constrained QR. We will estimate 19 equidistant quantiles for all the estimators, i.e. every $5^{th}$ quantile.

\subsection{In-sample fit}

\begin{figure}
    \centering
    \includegraphics[width=\linewidth]{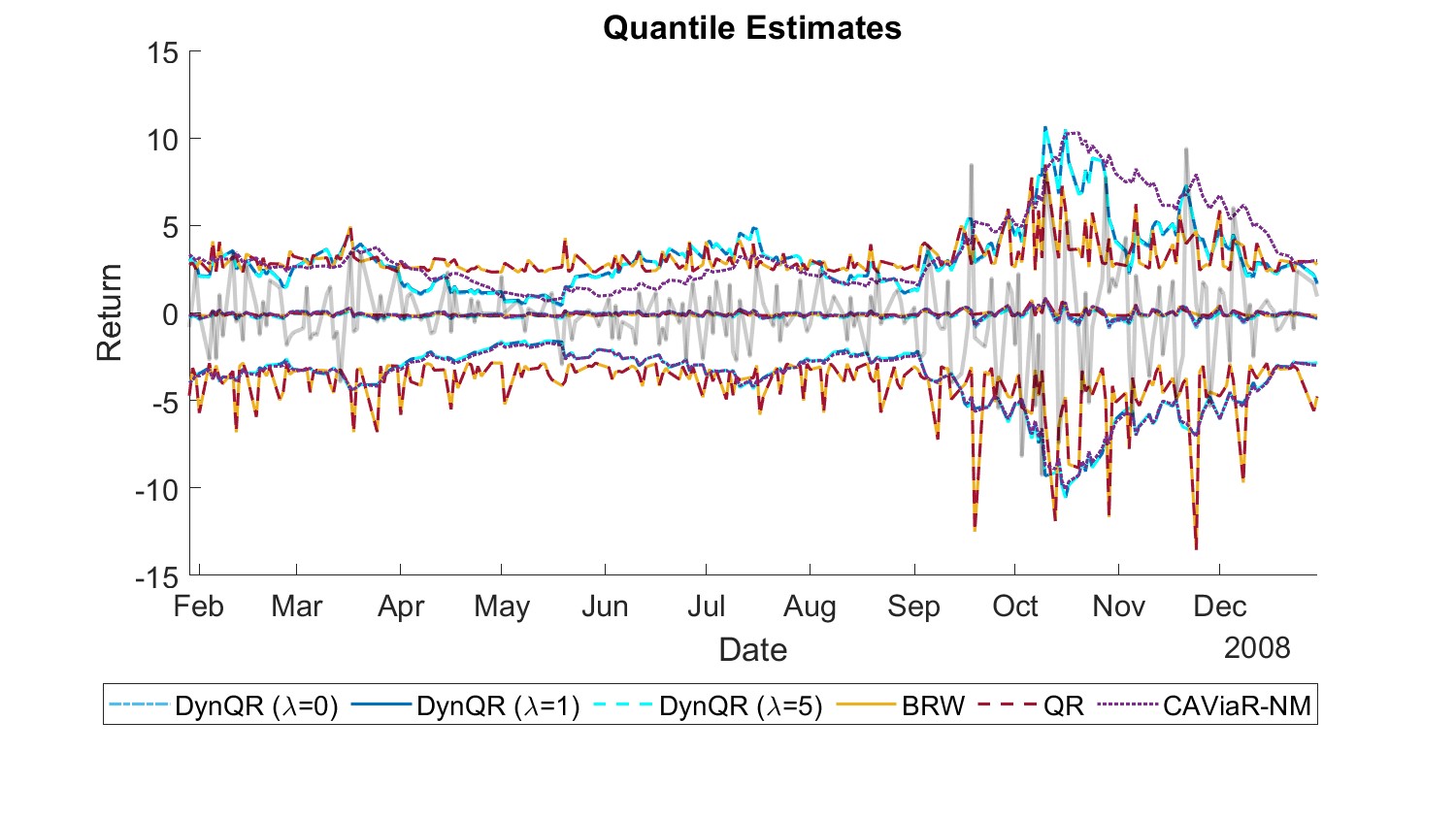}
    \caption{Fitted quantiles of the different estimators and observed returns}
    \label{fig:FitQuant}
\end{figure}

Figure (\ref{fig:FitQuant}) shows the selected fitted quantiles of the various estimated models. In particular, we show the $5^{th}$, $50^{th}$, and $95^{th}$ quantiles along with the observed returns.

Around October and November 2008, during the height of the financial crisis, there is a noticeable divergence among the quantile estimates. The $95^{th}$ and $5^{th}$ quantiles of estimators that allow for lagged quantiles appear to capture the spike in volatility more effectively. In particular, we can see the tails widening and narrowing over time, providing a clearer depiction of risks over time. In contrast, QR and BRW struggle to keep up, with their estimates showing abrupt shifts that may overstate or misrepresent changes in volatility. While the $50^{th}$ quantile is relatively more stable, during October and November, there is noticeable day-to-day fluctuation for all the estimators.

The quantiles of the DynQR and CAViaR-NM react strongly yet smoothly to extreme market movements, such as the large returns spikes seen in late 2008. This contrasts with the QR and BRW estimates, which can ``overshoot'' or ``undershoot'' during these periods. The dynamic quantile estimators have substantial and persistent variation in volatility, as reflected by the changing gap between the 5th and 95th quantiles throughout the sample period. This contrasts sharply with traditional QR methods (QR and BRW), whose estimated quantiles portray more volatility, but the overall distribution does not yield a persistent gap between the extreme quantiles during the financial crisis. In this way the the traditional quantiles are more volatile without resulting in a persistent increase in the volatility of the distribution. The smoothness of the dynamic quantile models likely results from the inclusion of lagged quantiles, which help stabilise the estimates over time.

For lower quantiles, the differences between CAViaR-NM and DynQR are minimal, with both methods producing similar fits. However, for upper quantiles, the divergence between these methods becomes more pronounced, particularly during the start of the financial crisis. This indicates that the modelling of upper-tail risks may vary substantially depending on the estimator employed. Note that the only difference between DynQR ($\lambda=0$) and CAViaR-NM is the optimisation routine and not the objective function.

\begin{table}[]
\centering
%\resizebox{\textwidth}{!}{%
\begin{tabular}{l|c}
\hline
\textbf{Estimator} & \textbf{Average Crossing Incidence} \\ \hline
DynQR ($\lambda=0$) & 0.113 \\
DynQR ($\lambda=1$) & 0.011 \\
DynQR ($\lambda=5$) & 0.004 \\
QR & 0.016 \\
CAViaR-NM & 0.316 \\ \hline
\end{tabular}%
%}
\caption{Quantile crossing incidence for the different estimators.}
\label{tab:Crossing}
\end{table}

The DynQR estimators with different penalty parameters show similar values for the quantiles presented. As such, the penalty for quantile crossing has minimal impact on these particular quantiles. However, this is not the case for the other quantiles. This can be seen from the average crossing incidence values of table (\ref{tab:Crossing}). Note that the BRW estimator enforces non-crossing quantiles in-sample and as such it's crossing incidence is omitted from this table.

The table corroborates the findings of the monte carlo section, that estimators that allow for lagged quantiles are more likely to have a higher degree of quantile crossing. In particular, the CAViar-NM and the DynQR with $\lambda=0$ both portray more crossing quantiles than the traditional quantile regression. As such, imposing some form of crossing penalty is necessary if one wants to estimate various quantiles with lagged quantile dynamics.

From the table we can see that as $\lambda>0$, the crossing incidence decreases. However, there are diminishing returns to this decrease. In particular, the bulk of the crossing incidence is diminished when setting $\lambda=1$, which corroborates the results of the Monte Carlo exercise. 

Perhaps the most striking finding of table (\ref{tab:Crossing}) is the large difference in crossing incidence between DynQR ($\lambda=0)$, with 11.3\% of quantiles crossing, and CAViar-NM, with 31.6\% of crossing quantiles. This again showcases the advantage of using CMA-ES algorithm instead of the Nelder-Mead algorithm.

\subsection{Coefficient profiles}
\subsubsection{Full Sample}
\begin{figure}
    \centering
    \includegraphics[width=\linewidth]{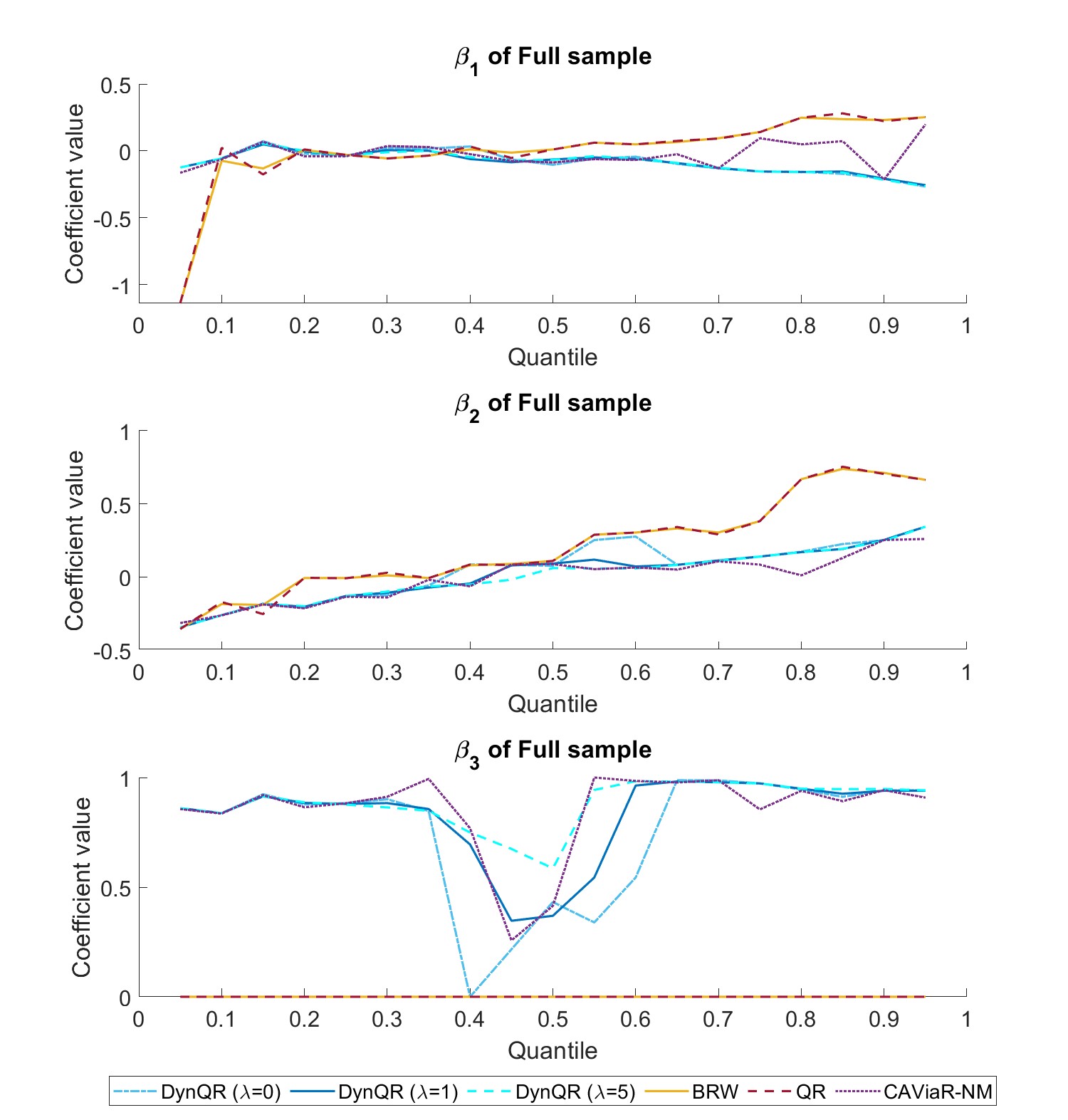}
    \caption{Coefficient profiles of the different estimators}
    \label{fig:Coeffs}
\end{figure}

As seen in the coefficient profiles in figure (\ref{fig:Coeffs}) the different penalty terms yield differences in coefficients at intermediate quantiles. In this figure $\beta_1$ is the impact of positive past deviations ($|y_{t-1}^T|^{+}$), $\beta_2$ is the impact of negative past deviations ($|y_{t-1}^T|^{-}$), and $\beta_3$ is the impact of past lagged quantiles.

Across the quantile range, $\beta_1$ demonstrates relatively stable coefficients across quantiles for all methods, with an overall downward slope for the lagged quantile estimators. This means that positive past deviations reduce future volatility for these models. DynQR with different penalty parameters are highly consistent, showing minimal divergence across quantiles, reflecting that penalizing quantile crossing has little influence on the coefficient for positive past deviations. Interestingly, the CAViaR-NM estimator showcases more ``jagged'' profiles for the coefficient for the upper quantiles. This is in line with the finding of \citet{szendrei2023fused}, who show that non-crossing constraints help limit variation in the coefficient profile across quantiles. Relative to these estiamtors, the QR and BRW models display more fluctuation, at the lower quantiles (below 0.3). At these lower quantiles, the traditional quantile estimators show more variation in the coefficient across quantiles. Furthermore, the shape of the coefficient shows an opposite profile, i.e. upwards sloping, for the traditional quantile estimators.

The coefficient $\beta_2$ grows gradually across the quantiles for all the estimators considered. This means that a negative return increases future volatility. For intermediate quantiles (0.3 to 0.7), DynQR estimators with different penalty parameters begin to diverge slightly from each other, indicating that the choice of $\lambda$ impacts the sensitivity to negative past deviations in intermediate, but not extreme, quantiles. QR and BRW exhibit relatively sharper transitions and less stability compared to DynQR. CAViaR-NM closely tracks DynQR estimators at lower quantiles but diverges slightly at higher quantiles.

The most striking variation is observed in $\beta_3$. DynQR estimators show considerable differences depending on the penalty term. Importantly, just like for $\beta_2$, these differences occur at intermediate quantiles only. Furthermore, CAViaR-NM aligns with DynQR for the extreme quantiles but diverges at central quantiles. These findings suggest that crossing is more likely to occur at these quantiles.

\newpage
\subsubsection{Rolling Sample}
\begin{figure}
    \centering
    \includegraphics[width=\linewidth]{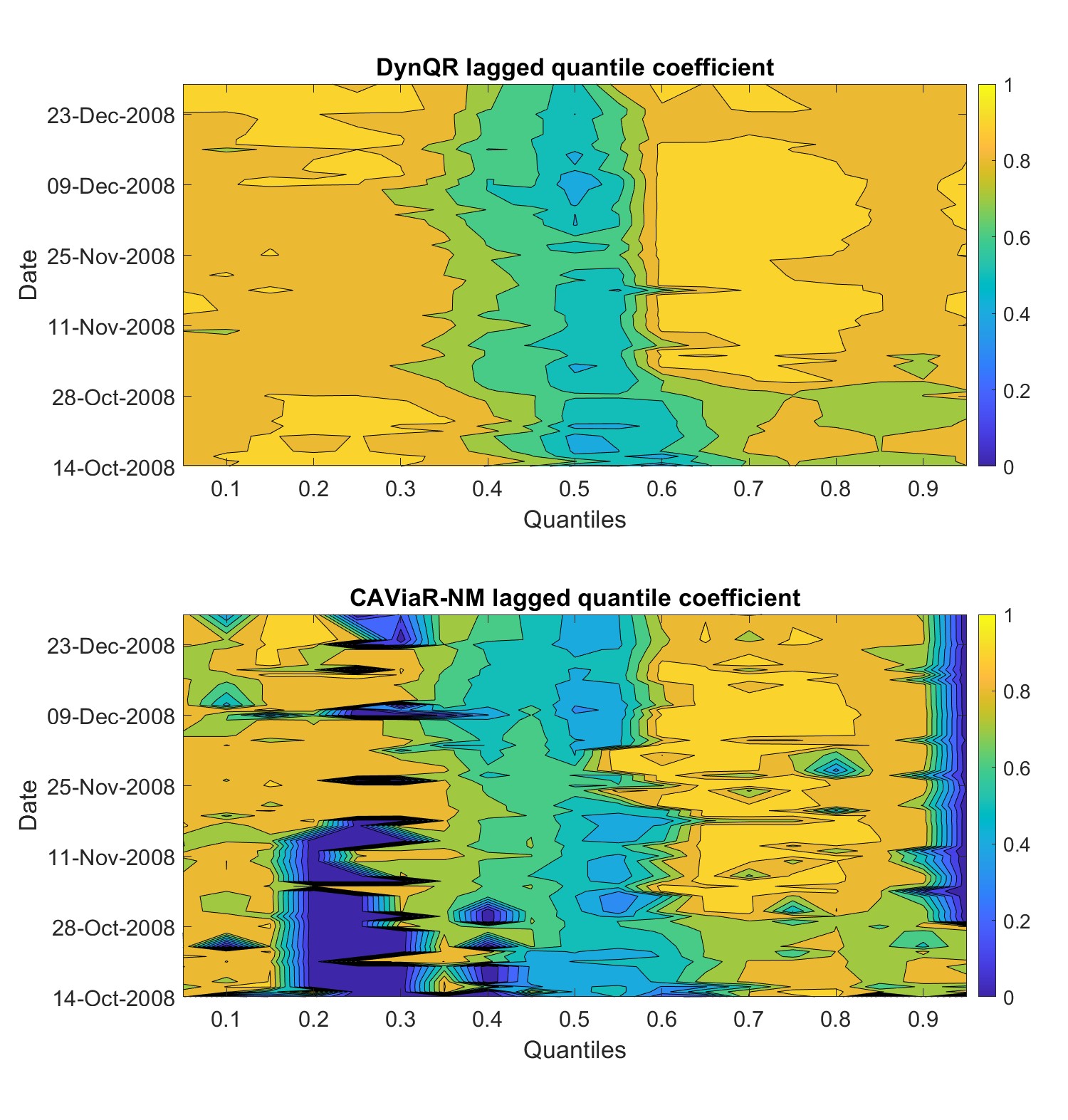}
    \caption{Coefficient profiles for lagged quantile on rolling samples}
    \label{fig:RollCoeffs}
\end{figure}

\citet{gourieroux2008dynamic} find some time variation in their estimated parameters for dynamic quantiles. To this end, we check for the presence of time variation. To do this we fit the model on a rolling window of 200 observations. In this way we will see how the coefficients evolve from 2008 October onwards. In the interest of space we will only present the lagged quantile coefficients of DynQR ($\lambda=1$) and CAViaR-NM. The contour plots of these coefficients (across time and across quantiles) is shown in figure (\ref{fig:RollCoeffs}).

The figure reveals that the coefficients of the DynQR are very stable over time. In particular, we observe a similar profile across quantiles for each window as we saw in the coefficients run on the full sample. The same cannot be said for the CAViaR-NM results, which showcase just as big variation across time as quantiles. Importantly, the quantile profile of the coefficients for the CAViaR-NM estimator changes drastically from one time period to the next, and at some periods (such as through October) does not resemble the quantile coefficient profile from the full sample.

Given the results from the Monte Carlo section, the lack of stability in the CAViaR-NM's quantile coefficient profiles is likely on account of data scarcity. Imposing non-crossing penalties (and constraints) helps the most when the number of observations is limited. As such, it is not surprising that the DynQR with $\lambda=1$ portrays more stable coefficients across time than the CAViaR-NM.

\subsection{Forecast performance}
\subsubsection{Main specification}
To evaluate the forecast performance of the different estimators, the quantile weighted CRPS (qwCRPS) of \citet{gneiting2011comparing} is chosen as a scoring rule. To calculate this measure, we first take the Quantile Score (QS), which is the weighted residual for a given forecast observation, $\hat{y}_{t+h,\tau_q}$. Using the QS the qwCRPS is calculated as:

\begin{equation}
    qwCRPS_{t+h} = \int^1_0 \; w_{\tau_q} QS_{t+h,\tau_q}d\tau,
\end{equation}

where $w_{\tau_q}$ denotes a weighting scheme to evaluate specific parts of the forecast density. The choice of this measure as the scoring rule enables us to evaluate differences at different part of the distribution, by applying different weighting schemes. We consider four such weighting schemes: $w_{\tau_q}^1=\frac{1}{Q}$ places equal weight on all quantiles;\footnote{This is equivalent to taking the average of the weighted residuals at a given observation.} $w_{\tau_q}^2=q(1-\tau_q)$ places more weight on central quantiles; $w_{\tau_q}^3=(1-\tau_q)^2$ places more weight on the left tail; and $w_{\tau_q}^4=\tau_q^2$ more weight on the right tail. 

We will also produce the same scoring rule for the sorted forecasted quantiles, where we apply the method of \citet{chernozhukov2010quantile} to get monotonically increasing forecasted quantile estimates. We note that increasing $\lambda$ will only ensure non-crossing in-sample, while in \citet{szendrei2023fused} the hyperparameter values larger than 1 lead to less out-of-sample crossing. This is because in the adaptive non-crossing constraints setting, a hyperparameter value above 1 will penalise quantile closeness as well as quantile crossing.\footnote{If one would want to reduce out-of-sample crossing, equation \ref{eq:DynQR} needs to be changed so that it penalises in-sample quantile closeness as well. This can be implemented by setting $\varphi>0$ in $\min(0,[\hat{\mathcal{Q}}_{\tau_q,t-1}-\hat{\mathcal{Q}}_{\tau_{q-1},t-1}]-\varphi)$. We leave for future research how this change impacts the forecast performance.} The results using the different weighting schemes are presented in Table (\ref{tab:forcresults}).

The forecast results show that the estimators with quantile dynamics yield lower forecast errors for QS. In particular, we can see that the proposed DynQR estimators yield the lowest forecast errors for all parts of the density. Just like in the Monte Carlo exercise, we find the $\lambda=5$ leads to the best performance overall, but not much better than $\lambda=1$. We can also see from the table that the DynQR estimator has the biggest improvement over traditional quantile estimators at the tails, highlighting how volatility (and potentially skewness) in the FTSE100 had inertia. Interestingly for the unsorted forecasted quantiles, the CAViaR-NM estimator does not yield better central quantiles than the QR and BRW, and it only beats the QR for the left tail. The CAViaR-NM only yields better results than the traditional QR estimators at the right tail.

When sorting the forecasted quantiles, we can see that the DynQR still remains the best performer. Importantly, we see limited improvement for the traditional quantile estimators. This highlights how correcting for quantile crossing can only lead to improvements if the model is not misspecified. Of all the estimators, sorting helps CAViaR-NM and DynQR with $\lambda=0$ the most, corroborating our findings from before, that dynamic quantiles can lead to more quantiles crossing. Furthermore, we see no improvement in forecast results from sorting the other DynQR estimators.

\begin{table}[]
\centering 
\resizebox{\textwidth}{!}{%
\begin{tabular}{l|cccc|cccc} 
\hline 
& \multicolumn{4}{c|}{Unsorted} & \multicolumn{4}{c}{Sorted} \\ 
\hline 
& QS & Centre & Left tail & Right tail & QS & Centre & Left tail & Right tail 
\\ 
\hline 
DynQR $(\lambda=0)$ & 0.761 & 0.148 & 0.240 & 0.372 & 0.759 & 0.147 & 0.240 & 0.371 \\ 
DynQR $(\lambda=1)$ & 0.754 & 0.147 & 0.237 & 0.370 & 0.754 & 0.147 & 0.237 & 0.370 \\ 
DynQR $(\lambda=5)$ & 0.753 & 0.146 & 0.237 & 0.369 & 0.753 & 0.146 & 0.237 & 0.369 \\
BRW & 0.776 & 0.149 & 0.244 & 0.383 & 0.776 & 0.149 & 0.244 & 0.383 \\
QR & 0.777 & 0.149 & 0.245 & 0.383 & 0.776 & 0.149 & 0.244 & 0.383 \\
CAViaR-NM & 0.770 & 0.149 & 0.244 & 0.377 & 0.765 & 0.148 & 0.243 & 0.374 \\
\hline 
\end{tabular}
}%
\caption{Forecast results of the different estimators}
\label{tab:forcresults}
\end{table}

\subsubsection{Alternative specification}
While the results of the DynQR are reassuring, it can be argued that the traditional QR estimators are at a disadvantage compared to the CAViaR-NM and DynQR. In particular, the addition of the lagged quantile variable might track system-wide financial stress. To this end we follow the growth-at-risk literature \citep{adrian2019vulnerable,figueres2020vulnerable,szendrei2023revisiting} and include an additional exogenous variable that tracks financial stress. We will use the daily 1 factor version of \citet{varga2024non} which has been shown to have good short forecast-horizon properties, when the interest is in capturing the left tail of GDP growth distribution of the UK. The forecast results for this specification are shown in Table (\ref{tab:forcresults_UKFSI}).

\begin{table}[t]
\centering 
\resizebox{\textwidth}{!}{%
\begin{tabular}{l|cccc|cccc} 
\hline 
& \multicolumn{4}{c|}{Unsorted} & \multicolumn{4}{c}{Sorted} \\ 
\hline 
& QS & Centre & Left tail & Right tail & QS & Centre & Left tail & Right tail 
\\ 
\hline 
DynQR $(\lambda=0)$ & 0.773 & 0.150 & 0.248 & 0.375 & 0.766 & 0.149 & 0.245 & 0.373 \\ 
DynQR $(\lambda=1)$ & 0.770 & 0.149 & 0.247 & 0.373 & 0.768 & 0.149 & 0.247 & 0.372 \\ 
DynQR $(\lambda=5)$ & 0.763 & 0.148 & 0.245 & 0.371 & 0.762 & 0.148 & 0.244 & 0.370 \\
BRW                 & 0.771 & 0.149 & 0.244 & 0.377 & 0.771 & 0.149 & 0.244 & 0.377 \\
QR                  & 0.775 & 0.150 & 0.246 & 0.379 & 0.773 & 0.149 & 0.246 & 0.378 \\
CAViaR-NM           & 0.771 & 0.149 & 0.246 & 0.376 & 0.767 & 0.148 & 0.245 & 0.374 \\
\hline 
\end{tabular}
}%
\caption{Forecast results of the different estimators (Alternative specification)}
\label{tab:forcresults_UKFSI}
\end{table}

The table reveals that including a measure of financial stress helps the traditional quantile estimators, as their overall forecast performance improved. In particular, inclusion of a measure of financial stress helps in estimating the right side of the distribution for all estimators, while it yields improvements across the distribution for the QR and BRW. Nevertheless, the DynQR estimator with $\lambda>0$ still yields better overall forecast performance, regardless of including the additional variable. 

Interestingly, the forecast performance of DynQR is worse when including a measure of financial stress as can be seen in tables (\ref{tab:forcresults}) and (\ref{tab:forcresults_UKFSI}). This is on account of no shrinkage being included in the model and as such the influence of variables that only introduce noise in the model are not shrunk away towards zero. This in turn leads to worse forecast performance. As such, the lagged coefficient is not simply tracking system-wide financial stress. On account of this, a specification with lagged quantiles and no exogenous financial stress variable is preffered to capture the value-at-risk of the FTSE100.

When comparing the different $\lambda$ values for DynQR, we see a similar profile as before, namely that increasing the crossing penalty leads to better forecast performance. However, now the improvement from $\lambda=1$ to $\lambda=5$ is not only marginal. Furthermore, we can see in table (\ref{tab:forcresults_UKFSI}) sorting the forecasted quantiles leads to larger improvements for DynQR than previously. These findings indicate that crossing is more likely to occur as more variables are included. Given the equivalence between fused shrinkage and non-crossing \citep{szendrei2023fused}, this is not surprising as the inclusion of more variables leads to more fused variation, which in turn can lead to larger incidence of quantile crossing.

%Maybe also talk about rolling window forecast?
%Maybe also talk about non-asym slope specificaiton: show the coeffs and the forecast tables briefly?

\section{Conclusion}
This paper explored ways to impose non-crossing for the CAViaR process of \citet{engle2004caviar}. In particular, the paper derived a measure of average crossing distance that is on the same unit as the objective function. By doing so we propose a penalised regression framework, where the multi-quantile CAViaR process is penalised for crossing. Due to the ``rugged'' nature of the objective function, local optimisers such as the Nelder-Mead algorithm encounters difficulties in obtaining a satisfactory solution. To address this, we propose using the CMA-ES solver to find a solution for our crossing penalised CAViaR.

Through a Monte Carlo study, we find that the proposed estimator is capable of getting solutions extremely close to the traditional QR in QAR(1) settings when $\lambda=0$, and even reduce coefficient bias by increasing the penalty on crossing. Furthermore, the proposed optimisation algorithm is not influenced by the starting point for the iterations. Importantly, our method beats the traditional CAViaR proposed by \citet{engle2004caviar} and \citet{white2010modeling}.

An application to the FTSE100 during 2008 (which is a period characterised by high volatility) revealed that models incorporating lagged quantiles outperform traditional QR models in terms of capturing the evolving volatility, especially during extreme market events. Additionally, the proposed DynQR estimator with $\lambda>0$ provides superior forecast accuracy, particularly for the tails of the distribution, while minimizing quantile crossing. In contrast, traditional quantile estimators struggle to track the changing risk profile, particularly during the financial crisis. The comparison of in-sample fits, coefficient profiles, and out-of-sample forecast performance suggests that dynamic quantile models offer a more robust and stable approach to modelling financial market risks.

%An application to US GaR with the proposed DynQR (crossing penalised CAViaR with CMA-ES algorithm) also showed that the upper quantiles of GDP growth are characterised by ``sticky'' quantiles. This is evident from the estimated coefficient profiles, and supported by the forecasting performance of the DynQR. We find that setting $\lambda=1$ is sufficient to obtain satisfactory results and advise researchers to use this value when fitting a dynamic GaR.

In this paper, we only focused on pre-specified values for $\lambda$, which somewhat limits the performance of the method. Since the choice of $\lambda$ provides a way to control the penalty for quantile crossing, using a limited number of candidate values constrains the model's ability to adapt optimally to the data. On account of this, it is possible to achieve further gains with the crossing penalised CAViaR model by exploring a broader range of candidate hyperparameter values. However, this avenue comes at great computational burden as highlighted by the time required for each run in the Monte Carlo study. As such, future research could focus on developing hyperparameter tuning frameworks for the crossing penalised CAViaR model, that balance the trade-off between computation time and improved model's performance.

Building on the results of \citet{szendrei2023fused}, where penalising beyond quantile crossing led to better out-of-sample forecast results, another promising direction for future research is exploring the potential to penalise quantile closeness. As mentioned earlier, this could be achieved by setting $\varphi>0$ in $\min(0, [\hat{\mathcal{Q}}_{\tau_q,t-1} - \hat{\mathcal{Q}}_{\tau_{q-1},t-1}] - \varphi)$. While conceptually simple, the difficulty lies in its estimation, given that this change necessitates selecting two hyperparameters: $\lambda$ to penalise quantile crossing and $\varphi$ to shift close quantiles towards `pseudo-crossing', i.e quantile differences at time $t$ that are penalised only because they are pushed below the thresholding value by $\varphi$.

\pagebreak

%%TC:ignore
\bibliographystyle{chicago}
%\addcontentsline{toc}{chapter}{Bibliography}
%\pagestyle{ref}
\bibliography{main.bbl}

\begin{thebibliography}{}

\bibitem[\protect\citeauthoryear{Adrian, Boyarchenko, and Giannone}{Adrian et~al.}{2019}]{adrian2019vulnerable}
Adrian, T., N.~Boyarchenko, and D.~Giannone (2019).
\newblock Vulnerable growth.
\newblock {\em American Economic Review\/}~{\em 109\/}(4), 1263--1289.

\bibitem[\protect\citeauthoryear{Audet and Hare}{Audet and Hare}{2017}]{audet2017derivative}
Audet, C. and W.~Hare (2017).
\newblock Derivative-free and blackbox optimization.

\bibitem[\protect\citeauthoryear{Bollerslev}{Bollerslev}{1986}]{bollerslev1986generalized}
Bollerslev, T. (1986).
\newblock Generalized autoregressive conditional heteroskedasticity.
\newblock {\em Journal of econometrics\/}~{\em 31\/}(3), 307--327.

\bibitem[\protect\citeauthoryear{Bondell, Reich, and Wang}{Bondell et~al.}{2010}]{bondell2010noncrossing}
Bondell, H.~D., B.~J. Reich, and H.~Wang (2010).
\newblock Noncrossing quantile regression curve estimation.
\newblock {\em Biometrika\/}~{\em 97\/}(4), 825--838.

\bibitem[\protect\citeauthoryear{Boyd and Vandenberghe}{Boyd and Vandenberghe}{2004}]{boyd2004convex}
Boyd, S. and L.~Vandenberghe (2004).
\newblock {\em Convex optimization}.
\newblock Cambridge university press.

\bibitem[\protect\citeauthoryear{Chen, Gerlach, Hwang, and McAleer}{Chen et~al.}{2012}]{chen2012forecasting}
Chen, C.~W., R.~Gerlach, B.~B. Hwang, and M.~McAleer (2012).
\newblock Forecasting value-at-risk using nonlinear regression quantiles and the intra-day range.
\newblock {\em International Journal of Forecasting\/}~{\em 28\/}(3), 557--574.

\bibitem[\protect\citeauthoryear{Chernozhukov, Fern{\'a}ndez-Val, and Galichon}{Chernozhukov et~al.}{2010}]{chernozhukov2010quantile}
Chernozhukov, V., I.~Fern{\'a}ndez-Val, and A.~Galichon (2010).
\newblock Quantile and probability curves without crossing.
\newblock {\em Econometrica\/}~{\em 78\/}(3), 1093--1125.

\bibitem[\protect\citeauthoryear{Chernozhukov and Hong}{Chernozhukov and Hong}{2003}]{chernozhukov2003mcmc}
Chernozhukov, V. and H.~Hong (2003).
\newblock An mcmc approach to classical estimation.
\newblock {\em Journal of econometrics\/}~{\em 115\/}(2), 293--346.

\bibitem[\protect\citeauthoryear{Chernozhukov and Umantsev}{Chernozhukov and Umantsev}{2001}]{chernozhukov2001conditional}
Chernozhukov, V. and L.~Umantsev (2001).
\newblock Conditional value-at-risk: Aspects of modeling and estimation.
\newblock {\em Empirical Economics\/}~{\em 26}, 271--292.

\bibitem[\protect\citeauthoryear{D'Errico}{D'Errico}{2023}]{fminsearchbnd}
D'Errico, J. (2023).
\newblock fminsearchbnd, fminsearchcon.
\newblock MATLAB File Exchange, https://www.mathworks.com/matlabcentral/fileexchange/8277-fminsearchbnd-fminsearchcon. Retrieved: 10.02.2023.

\bibitem[\protect\citeauthoryear{Engle and Manganelli}{Engle and Manganelli}{2004}]{engle2004caviar}
Engle, R.~F. and S.~Manganelli (2004).
\newblock Caviar: Conditional autoregressive value at risk by regression quantiles.
\newblock {\em Journal of business \& economic statistics\/}~{\em 22\/}(4), 367--381.

\bibitem[\protect\citeauthoryear{Figueres and Jaroci{\'n}ski}{Figueres and Jaroci{\'n}ski}{2020}]{figueres2020vulnerable}
Figueres, J.~M. and M.~Jaroci{\'n}ski (2020).
\newblock Vulnerable growth in the euro area: Measuring the financial conditions.
\newblock {\em Economics Letters\/}~{\em 191}, 109126.

\bibitem[\protect\citeauthoryear{Galvao~Jr, Montes-Rojas, and Olmo}{Galvao~Jr et~al.}{2011}]{galvao2011threshold}
Galvao~Jr, A.~F., G.~Montes-Rojas, and J.~Olmo (2011).
\newblock Threshold quantile autoregressive models.
\newblock {\em Journal of Time Series Analysis\/}~{\em 32\/}(3), 253--267.

\bibitem[\protect\citeauthoryear{Gneiting and Ranjan}{Gneiting and Ranjan}{2011}]{gneiting2011comparing}
Gneiting, T. and R.~Ranjan (2011).
\newblock Comparing density forecasts using threshold-and quantile-weighted scoring rules.
\newblock {\em Journal of Business \& Economic Statistics\/}~{\em 29\/}(3), 411--422.

\bibitem[\protect\citeauthoryear{Gouri{\'e}roux and Jasiak}{Gouri{\'e}roux and Jasiak}{2008}]{gourieroux2008dynamic}
Gouri{\'e}roux, C. and J.~Jasiak (2008).
\newblock Dynamic quantile models.
\newblock {\em Journal of econometrics\/}~{\em 147\/}(1), 198--205.

\bibitem[\protect\citeauthoryear{Hansen}{Hansen}{2006}]{hansen2006cma}
Hansen, N. (2006).
\newblock The cma evolution strategy: a comparing review.
\newblock {\em Towards a new evolutionary computation: Advances in the estimation of distribution algorithms\/}, 75--102.

\bibitem[\protect\citeauthoryear{Hansen}{Hansen}{2009}]{hansen2009benchmarking}
Hansen, N. (2009).
\newblock Benchmarking a bi-population cma-es on the bbob-2009 function testbed.
\newblock In {\em Proceedings of the 11th annual conference companion on genetic and evolutionary computation conference: late breaking papers}, pp.\  2389--2396.

\bibitem[\protect\citeauthoryear{Hansen}{Hansen}{2016}]{hansen2016cma}
Hansen, N. (2016).
\newblock The cma evolution strategy: A tutorial.
\newblock {\em arXiv preprint arXiv:1604.00772\/}.

\bibitem[\protect\citeauthoryear{Hansen and Kern}{Hansen and Kern}{2004}]{hansen2004evaluating}
Hansen, N. and S.~Kern (2004).
\newblock Evaluating the cma evolution strategy on multimodal test functions.
\newblock In {\em International conference on parallel problem solving from nature}, pp.\  282--291. Springer.

\bibitem[\protect\citeauthoryear{Hansen, Niederberger, Guzzella, and Koumoutsakos}{Hansen et~al.}{2009}]{hansen2009tec}
Hansen, N., S.~Niederberger, L.~Guzzella, and P.~Koumoutsakos (2009).
\newblock A method for handling uncertainty in evolutionary optimization with an application to feedback control of combustion.
\newblock {\em IEEE Transactions on Evolutionary Computation\/}~{\em 13\/}(1), 180--197.

\bibitem[\protect\citeauthoryear{Hansen and Ostermeier}{Hansen and Ostermeier}{2001}]{hansen2001completely}
Hansen, N. and A.~Ostermeier (2001).
\newblock Completely derandomized self-adaptation in evolution strategies.
\newblock {\em Evolutionary computation\/}~{\em 9\/}(2), 159--195.

\bibitem[\protect\citeauthoryear{He}{He}{1997}]{he1997quantile}
He, X. (1997).
\newblock Quantile curves without crossing.
\newblock {\em The American Statistician\/}~{\em 51\/}(2), 186--192.

\bibitem[\protect\citeauthoryear{Koenker and Bassett}{Koenker and Bassett}{1978}]{koenker1978regression}
Koenker, R. and G.~Bassett (1978).
\newblock Regression quantiles.
\newblock {\em Econometrica: journal of the Econometric Society\/}~{\em 46\/}(1), 33--50.

\bibitem[\protect\citeauthoryear{Koenker and Xiao}{Koenker and Xiao}{2006}]{koenker2006quantile}
Koenker, R. and Z.~Xiao (2006).
\newblock Quantile autoregression.
\newblock {\em Journal of the American statistical association\/}~{\em 101\/}(475), 980--990.

\bibitem[\protect\citeauthoryear{Koenker and Zhao}{Koenker and Zhao}{1996}]{koenker1996conditional}
Koenker, R. and Q.~Zhao (1996).
\newblock Conditional quantile estimation and inference for arch models.
\newblock {\em Econometric theory\/}~{\em 12\/}(5), 793--813.

\bibitem[\protect\citeauthoryear{Nelder and Mead}{Nelder and Mead}{1965}]{nelder1965simplex}
Nelder, J.~A. and R.~Mead (1965).
\newblock A simplex method for function minimization.
\newblock {\em The computer journal\/}~{\em 7\/}(4), 308--313.

\bibitem[\protect\citeauthoryear{Rubia and Sanchis-Marco}{Rubia and Sanchis-Marco}{2013}]{rubia2013downside}
Rubia, A. and L.~Sanchis-Marco (2013).
\newblock On downside risk predictability through liquidity and trading activity: A dynamic quantile approach.
\newblock {\em International Journal of Forecasting\/}~{\em 29\/}(1), 202--219.

\bibitem[\protect\citeauthoryear{Szendrei, Bhattacharjee, and Schaffer}{Szendrei et~al.}{2024}]{szendrei2023fused}
Szendrei, T., A.~Bhattacharjee, and M.~E. Schaffer (2024).
\newblock Fused {LASSO} as non-crossing quantile regression.
\newblock {\em arXiv preprint arXiv:2403.14036\/}.

\bibitem[\protect\citeauthoryear{Szendrei and Varga}{Szendrei and Varga}{2023}]{szendrei2023revisiting}
Szendrei, T. and K.~Varga (2023).
\newblock Revisiting vulnerable growth in the euro area: Identifying the role of financial conditions in the distribution.
\newblock {\em Economics Letters\/}, 110990.

\bibitem[\protect\citeauthoryear{Varga and Szendrei}{Varga and Szendrei}{2025}]{varga2024non}
Varga, K. and T.~Szendrei (2025).
\newblock Non-stationary financial risk factors and macroeconomic vulnerability for the uk.
\newblock {\em International Review of Financial Analysis\/}~{\em 97}, 103866.

\bibitem[\protect\citeauthoryear{White, Kim, and Manganelli}{White et~al.}{2010}]{white2010modeling}
White, H., T.-H. Kim, and S.~Manganelli (2010).
\newblock Modeling autoregressive conditional skewness and kurtosis with multi-quantile caviar.
\newblock In {\em Volatility and Time Series Econometrics: Essays in Honor of Robert Engle}. Oxford University Press.

\bibitem[\protect\citeauthoryear{White, Kim, and Manganelli}{White et~al.}{2015}]{white2015var}
White, H., T.-H. Kim, and S.~Manganelli (2015).
\newblock Var for var: Measuring tail dependence using multivariate regression quantiles.
\newblock {\em Journal of econometrics\/}~{\em 187\/}(1), 169--188.

\bibitem[\protect\citeauthoryear{Xiao and Koenker}{Xiao and Koenker}{2009}]{xiao2009conditional}
Xiao, Z. and R.~Koenker (2009).
\newblock Conditional quantile estimation for generalized autoregressive conditional heteroscedasticity models.
\newblock {\em Journal of the American Statistical Association\/}~{\em 104\/}(488), 1696--1712.

\end{thebibliography}
%%TC:endignore

\pagebreak

%\section{Graphs}
%\include{Chapters/appfiguresincl}

%\appendix 
%\section{Appendix}
%\input{Chapters/Appendix}

\end{document}